\documentclass[superscriptaddress,aps,pra,showkeys,10pt]{revtex4-2}
\usepackage{amsmath,mathtools,siunitx}
\usepackage{amssymb}
\usepackage{graphicx}
\usepackage{color}
\usepackage{bm}
\usepackage[colorlinks,linkcolor=blue,anchorcolor=blue,citecolor=blue,urlcolor=blue]{hyperref}
\usepackage{booktabs}
\usepackage{tabularx}

\begin{document}
\title{Entanglement enhancement of two giant atoms with multiple connection points in bidirectional-chiral quantum waveguide-QED system}

\author{Jie Liu}
\affiliation{Lanzhou Center for Theoretical Physics, Key Laboratory of Theoretical Physics of Gansu Province, and Key Laboratory of Quantum Theory and Applications of MoE, Lanzhou University, Lanzhou, Gansu 730000, China}

\author{Yue Cai}
\affiliation{Lanzhou Center for Theoretical Physics, Key Laboratory of Theoretical Physics of Gansu Province, and Key Laboratory of Quantum Theory and Applications of MoE, Lanzhou University, Lanzhou, Gansu 730000, China}

\author{Kang-Jie Ma}
\affiliation{Lanzhou Center for Theoretical Physics, Key Laboratory of Theoretical Physics of Gansu Province, and Key Laboratory of Quantum Theory and Applications of MoE, Lanzhou University, Lanzhou, Gansu 730000, China}

\author{Lei Tan}
\email{tanlei@lzu.edu.cn}
\affiliation{Lanzhou Center for Theoretical Physics, Key Laboratory of Theoretical Physics of Gansu Province, and Key Laboratory of Quantum Theory and Applications of MoE, Lanzhou University, Lanzhou, Gansu 730000, China}

\author{Wu-Ming Liu}
\affiliation{Beijing National Laboratory for Condensed Matter Physics, Institute of Physics, Chinese Academy of Sciences, Beijing 100190, China}

\begin{abstract}
We study the entanglement generation of two giant atoms within a one-dimensional bidirectional-chiral waveguide quantum electrodynamics (QED) system, where the initial state of the two giant atoms are $|e_a,g_b\rangle $. Here, each giant atom is coupled to the waveguide through three connection points, with the configurations divided into five types based on the arrangement of coupling points between the giant atoms and the waveguide: separate, fully braided, partially braided, fully nested, and partially nested. We explore the entanglement generation process within each configuration in both nonchiral and chiral coupling cases. It is demonstrated that entanglement can be controlled as needed by either adjusting the phase shift or selecting different configurations. For nonchiral coupling, the entanglement of each configuration exhibits steady state properties attributable to the presence of dark state. In addition, we find that steady-state entanglement can be obtained at more phase shifts  in certain configurations by increasing the number of coupling points between the giant atoms and the bidirectional waveguide. In the case of chiral coupling, the entanglement is maximally enhanced compared to the one of nonchiral case. Especially in fully braided configuration, the concurrence reaches its peak value 1, which is robust to chirality. We further show the influence of atomic initial states on the evolution of interatomic entanglement. Our scheme can be used for entanglement generation in chiral quantum networks of giant-atom waveguide-QED systems, with potential applications in quantum networks and quantum communications.

\keywords{Entanglement enhancement, Giant atom, Multiple connection points, Bidirectional-chiral waveguide}
\end{abstract}

\maketitle
\section{INTRODUCTION}

Quantum entanglement, as a manifestation of the principle of state superposition in multi-particle systems, is a vital resource in quantum communication and quantum networks \cite{Einstein1, Horodecki2}. In recent years, the generation of quantum entanglement has been both theoretically and experimentally explored across various physical systems \cite{Pan3, Leibfried4, Blinov5, Raimond6, Herbert7, Blais8, Blais9, Wallraff10}, among which waveguide-QED systems \cite{Sheremet22, Roy23}, characterized by strong coupling and one-dimensional confinement, provide an excellent platform for creating entanglements. In the system, quantum qubits can spontaneously create two-qubit entanglement via coupling to an infinite waveguide \cite{Gonzalez-Tudela24, Cano25, Carlos26, Gonzalez-Ballestero27, Paolo28, Kien29}.

Quantum systems interaction with light in waveguides usually exhibits equal coupling strengths in both left and right propagation directions. Notably, advances in chiral light-matter interface research have shown that this bidirectional symmetry can be broken \cite{Pichler30, Vermersch31, Yan32, Berman33}. This asymmetrical interaction, referred to as chiral coupling, is a direct consequence of optical spin–orbit coupling \cite{Bliokh34}. In recent years, chiral quantum networks have been extensively studied \cite{Pichler30, Odahl36, Ramos37, Mahmoodian38, Giuseppe39}, specifically, chiral waveguides have been demonstrated as a particularly suitable platform for enhancing maximum entanglement. It is shown that the maximum concurrence between two small atoms is 1.5 times that of the nonchiral system \cite{Gonzalez-Tudela24, Cano25, Gonzalez-Ballestero27, Paolo28, Gonzalez-Ballestero40, Zheng41, Mirza42}. Inspired by Ref. \cite{Gonzalez-Ballestero40}, Mok \emph{et al.} utilized directional asymmetry in chirally coupled single-mode ring resonators to generate entangled states between two small atoms \cite{Mok43}, achieving a maximum entanglement value 0.969. This represents a substantial enhancement compared to the value 0.736 reported in Ref. \cite{Gonzalez-Ballestero40}. Besides, the investigation of multiqubit entanglement of small atoms in bidirectional-chiral waveguides has also been undertaken \cite{Mirza45}. It shown that the introduction of chirality can increase the entanglement by more than a factor of 3/2 compared to the nonchiral case.
\begin{figure}[tbp]
	\centering
	\includegraphics[width=0.6\textwidth]{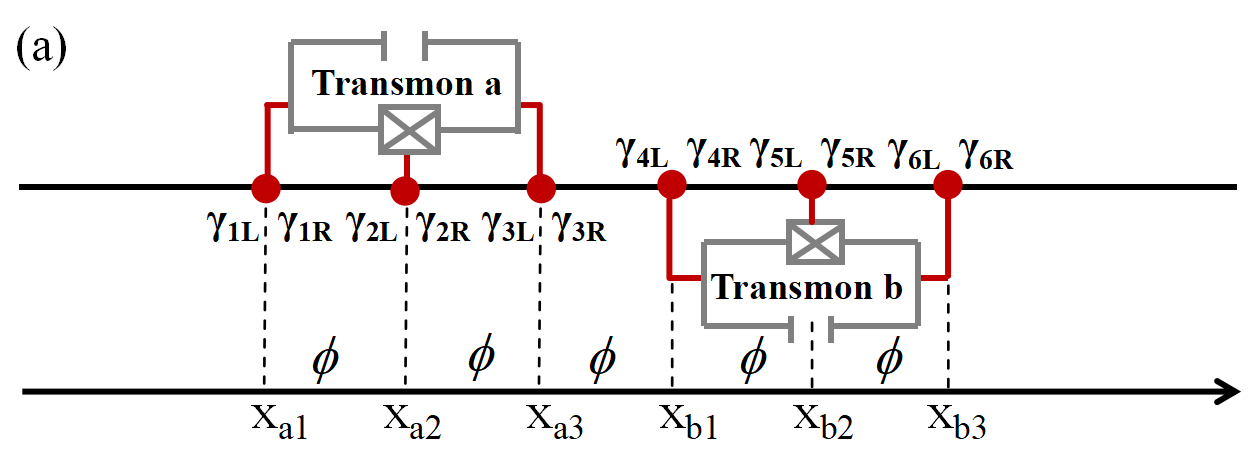}
	\newline\vspace{2pt}
	\includegraphics[width=0.6\textwidth]{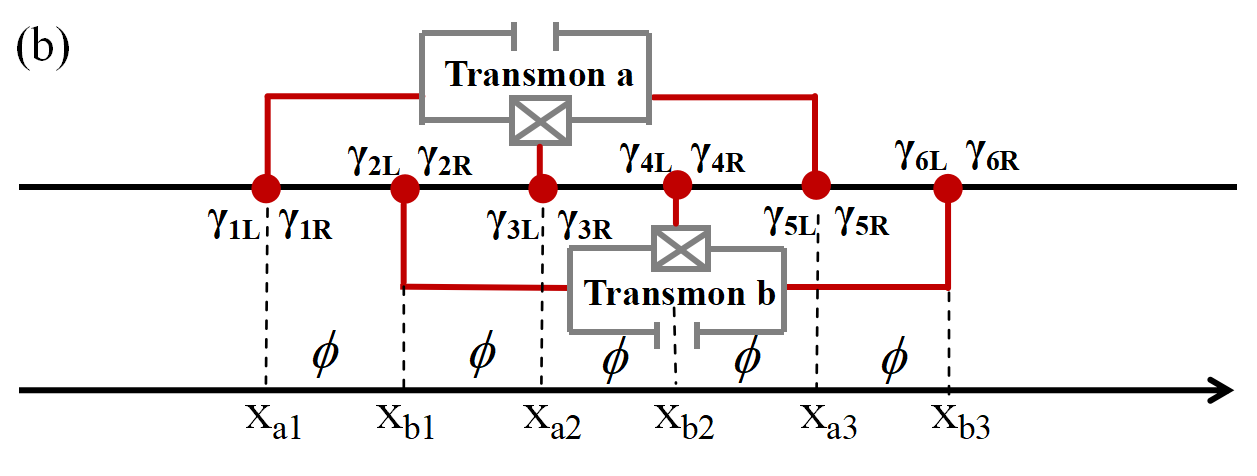}
	\newline\vspace{2pt}
	\includegraphics[width=0.6\textwidth]{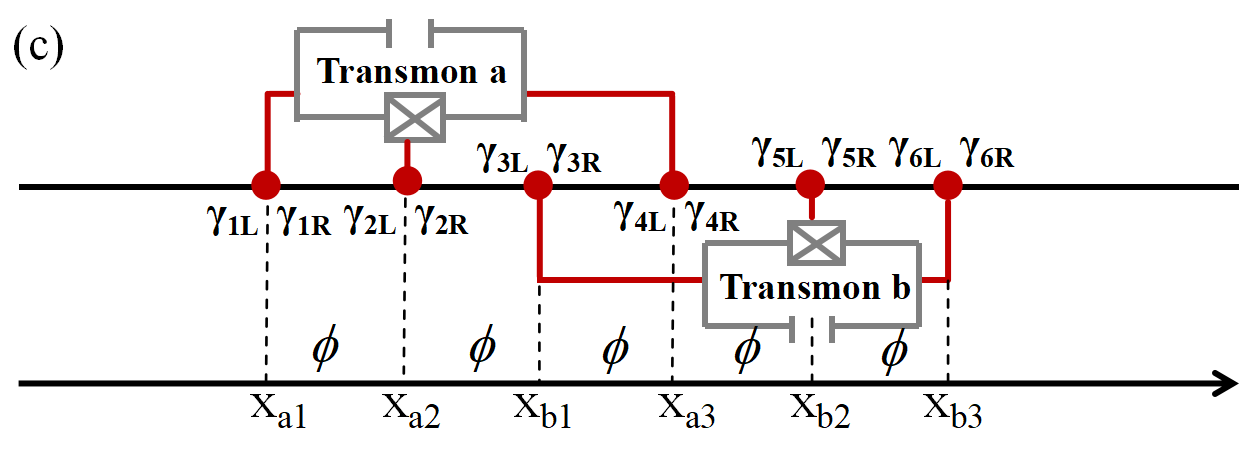}
	\newline\vspace{2pt}
	\includegraphics[width=0.6\textwidth]{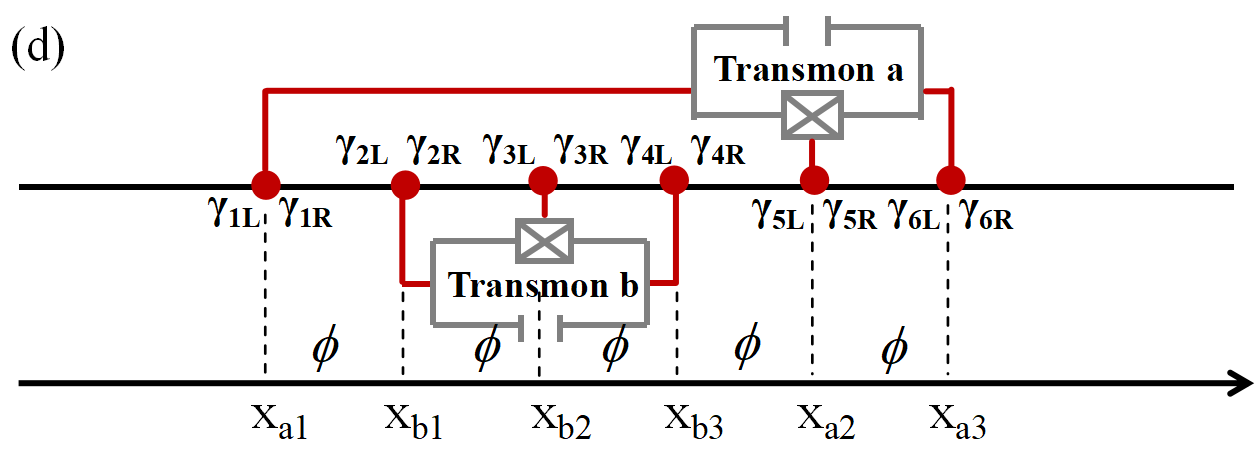}
	\newline\vspace{2pt}
	\hspace{-25mm}
	\includegraphics[width=0.6\textwidth]{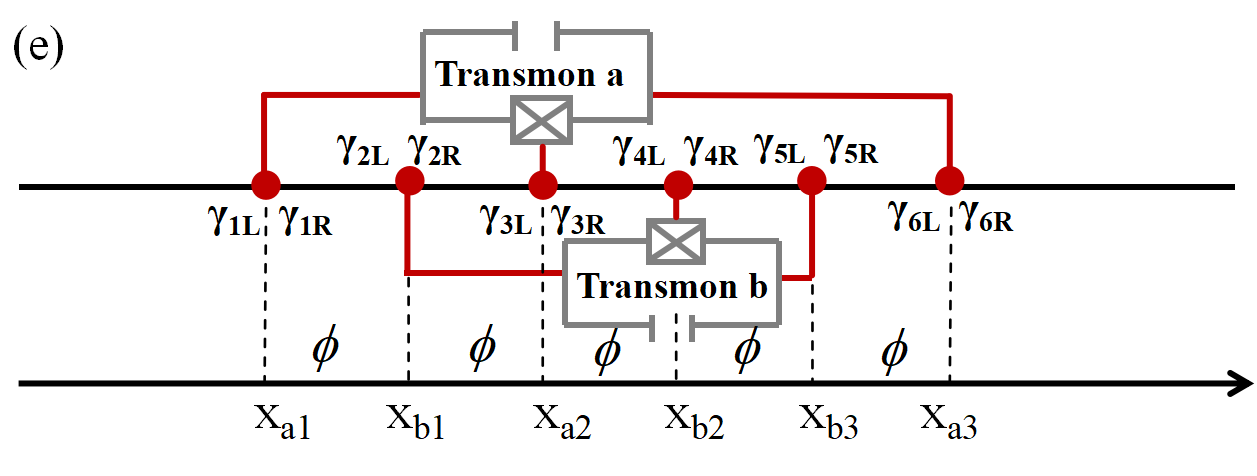}
	\caption{(Color online) Schematic of configurations for two giant atoms coupled to a bidirectional-chiral waveguide. (a) separated, (b) fully braided, (c) partially braided, (d) fully nested, (e) partially nested. The positions where the giant atoms couple to the waveguide are denoted by $x_{jn}$, where $j = a, b$ represent two giant atoms, and $n = 1, 2, 3$ correspond to the coupling points. For each coupling point, the strengths of the atom's coupling to the left and right propagating modes of the waveguide are denoted as $\gamma_{kR}$ and $\gamma_{kL}$, such that $\gamma_{kR} + \gamma_{kL} = \gamma_k$, where $k = 1,2,3,4,5,6$. In particular, in this paper we consider only the case of two-atom resonance, i.e. $\omega = \omega_a = \omega_b$. The phase shift between neighbouring coupling points is represented by $\phi = k_0d = \omega d / v_g$, which can be adjusted by tuning the frequency $\omega$ of the giant atom. Here, $k_0$ is the wave vector and $d$ is the fixed distance between the two connection points in experimental setups; $v_g$ represents the velocity of the modes in the waveguide. For convenience, in the calculations that follow, we assume that the coupling strength to the right-propagating mode of the waveguide at all points is $\gamma_{kR} \equiv \gamma_R$, and the same applies to all $\gamma_{kL}$. Therefore, $\gamma_k = \gamma = \gamma_{R} + \gamma_{L}$.}
	\label{fig1}
\end{figure}

The entanglement among atoms, as examined in the waveguide-QED systems above, is limited to small atoms. Recent years, an artificial atom known as a giant atom, distinguished by its large size, extends beyond the dipole approximation and is capable of establishing non-local coupling to the thermal reservoir. From the perspective of giant atoms, quantum control can be achieved by tuning the atomic coupling or size. This has attracted the attention of many researchers. Some work has been conducted on the giant-atom waveguide-QED systems, including the frequency-dependent relaxation \cite{Kockum46}, interatomic decoherence-free interactions \cite{Kockum47, Kannan48, Cilluffo49, Carollo50, Soro51}, non-Markovian electromagnetically induced transparency \cite{Zhu52}, and single-photon scattering \cite{Yin54, Zhao, Cai111, Li2024}. Specifically, all of the studies mentioned above on giant-atom waveguide-QED systems have involved multiple points coupling beyond two point coupling and yielded some interesting results.

For the entangled state of giant atoms, research has primarily focused on generating entanglement between two giant atoms coupled to a common waveguide at two connection points in nonchiral system \cite{Yin55, Yin56, 130, cai2024, Ma2024, Ma, luo57}. However, the entanglement generation of giant atoms with multiple coupling points in a waveguide-QED system has not yet been explored. To investigate the impact of quantum interference effects between multiple coupling points of giant atoms and the waveguide on entanglement dynamics, this paper extends the model proposed in Ref. \cite{Yin55} from two-point to three-point coupling and introduces chirality into the extended model. Here, the initial state of the two giant atoms are chosen as $|e_a,g_b\rangle $. Our study indicates that in chiral system, two giant atoms can exhibit greater entanglement compared to small atoms in certain configurations. In particular, the maximum entanglement value can reach 1 in fully braided configuration and this maximum value is independent of chirality, both of which are not possible in small-atom chiral system \cite{Gonzalez-Ballestero40}. In the nonchiral system, compared to the results of Ref. \cite{Yin55}, we find that an increase in the number of coupling points does not change the maximum entanglement but can produce steady-state entanglement over a greater range of phase shifts except for the fully braided configuration. Moreover, from the results of this paper on both nonchiral and chiral cases, the presence of chiral effects is beneficial for the entanglement enhancement between two giant atoms. We also explored the impact of initial atomic states on the entanglement dynamics between two giant atoms. The results demonstrate that entanglement dynamics vary for different single-excitation initial states under chiral coupling, independent of the permutation symmetry exhibited by the two giant atoms. In contrast, under nonchiral coupling, changes in entanglement dynamics are noted only when the giant atoms lack permutation symmetry.

The structure of this paper is outlined as follows. In Sec. \ref{II}, we describe the theoretical model of two giant atoms coupled to the commom one-dimensional (1D) bidirectional-chiral waveguide. We provide master equation and non-Hermitian Hamiltonian to describe the system's dynamic evolution after tracing out the waveguide mode, employing concurrence to characterize entanglement. In Sec. \ref{III}, we discuss the entanglement generated in both nonchiral and chiral cases under the single-excitation subspace, based on the configurations presented in Sec. \ref{II}. We also investigate the effects of phase shifts and coupling configurations on entanglement, and compare this to the entanglement in nonchiral case where two giant atoms are coupled to the waveguide at two points each. In Sec. \ref{IV}, we study the influence of atomic initial states on interatomic entanglement. In particular, by further studying fully braided configuration, we demonstrate in Sec. \ref{V} that the maximal entanglement is robust against the variations in chirality. Finally, we present a brief conclusion in Sec. \ref{VI}.

\section{MODEL AND THEORETICAL DESCRIPTION}\label{II}
The system considered in this paper is a giant-atom waveguide-QED system. In this system, two giant atoms, each with two levels, are coupled to a common open 1D waveguide. The giant atom is realized by coupling the two-level transmon qubit to a waveguide at three spatially separated points, as shown in Fig. \ref{fig1}. Based on the distinct arrangements of connection points when two giant atoms are coupled to the waveguide, three primary categories of coupling configurations can be identified: separated [Fig. \ref{fig1}(a)], fully braided [Fig. \ref{fig1}(b)], and fully nested [Fig. \ref{fig1}(d)]. Given the enhanced flexibility in connection points arrangements for giant atoms when coupled to the waveguide at three points each, this paper also considers two additional configurations, denoted respectively as partially braided [Fig. \ref{fig1}(c)] and partially nested [Fig. \ref{fig1}(e)]. In the text, we utilize the superscripts ``S'', ``FB'', ``PB'', ``FN'' and ``PN'' to denote separate-, fully braided-, partially braided-,  fully nested-, and partially nested configurations, respectively.

In general, the coupling of atoms to a reservoir (which we assume to be an open 1D waveguide) can be classified as either nonchiral (bidirectional) or chiral based on whether the symmetry of the waveguide propagation modes is broken. As depicted in Fig. \ref{fig1}, giant atoms are coupled to both the left and right modes of the waveguide at each connection point. When the coupling strength of giant atom to the left and right modes of the waveguide are symmetric, i.e., $\gamma_L / \gamma_R=1 $, we consider the giant atoms to be non-chirally coupled to the waveguide. However, when the symmetry of coupling strength is broken, i.e., $\gamma_L/\gamma_R \neq 1$, the giant atom is considered to be chirally coupled to the waveguide. Specifically, for the limit case of chiral coupling, the system is referred to as a cascade system (or a perfect chiral system) when we assume that the thermal reservoir excitations move only to the right (i.e., $\gamma_L = 0$), i.e., the giant atoms are only coupled to modes propagating in the right direction of the waveguide, where $R$ and $L$ denote left and right, respectively. The quantum channel of chiral coupling enables directional and selective qubit-qubit interactions. 

Because of the interaction between the giant atoms and the waveguide, the dissipative dynamics of the two giant atoms as an open subsystem can be described by the master equation. We assume that the field modes in the waveguide are initially in the vacuum state. In the Markov approximation (assuming that the time required for a photon to fly between neighbouring coupling points is much smaller than the lifetime of giant atoms and $\Gamma_j \ll \omega_j$, where $\omega_j$ is the eigenfrequency of the giant atoms), by tracing out the waveguide modes, employing the SLH formalism \cite{gough58, Gough59, combes60},  and working in the interaction picture, the general form of the Markovian quantum master equation of two giant atoms coupled to a bidirectional-chiral waveguide can be obtained \cite{Kockum47, Soro51}, where each giant atom is coupled to the waveguide at three location points. The SLH triplet consists of a scattering matrix $S$, a vector $L$ of $n$ collapse operators describing the coupling of the system to the waveguide, and a Hamiltonian $H$ for the system. The master equation is applicable to all setups considered in this paper.
\begin{align}
	\dot{\rho} = & -i \left[ \sum_{j=a,b} \delta\omega_j +  \left(g_{a,b} \sigma_{-}^{a}\sigma_{+}^{b} + \text{H.c.}\right), \rho \right] +\sum_{j=a,b} \Gamma_{j} \mathcal{D}[\sigma_{-}^{j}]\rho  \nonumber \\
	& + \sum_{j=a,b} \left[ \Gamma_{\text{coll},a,b} \sigma_{-}^{a}\rho\sigma_{+}^{b} - \frac{1}{2} \left\{ \sigma_{-}^{a}\sigma_{+}^{b}, \rho \right\}+ \text{H.c.}\right] 
	\label{eq:me}
\end{align}

The first term in Eq. (\ref{eq:me}) is the coherent term, representing the unity evolution of the system, including the interatomic interaction strength $g_{a,b}$ induced by virtual photons and the frequency shift $\delta\omega_j$ of giant atom $j$. The excitation and de-excitation processes of an atom $j$ are denoted by the raising and lowering operators $\sigma_+^j$ and $\sigma_-^j$ of the atom, respectively. The symbol $ H.c.$ denotes the Hermitian conjugate. The subsequent two terms correspond to the incoherent (dissipative) terms, contributing to the decay process. In which, $\Gamma_j$ and $\Gamma_{\text{coll},a,b}$ denote the individual decay rates of giant atom $j$ and collective decay rates of giant atom $a$ and $b$. The standard Lindblad superoperator describing the decay process is defined as $D[X]\rho = X\rho X^{\dagger} - \frac{1}{2} \{ X^{\dagger}X, \rho \}$, where $X$ can be an arbitrary operator. Here, we neglect dissipation arising from non-waveguide degrees of freedom, as it is significantly smaller compared to the decay rate of giant atoms in practical physical systems. It is important to note that when the phase shift difference between neighbouring coupling points is comparable to the wavelength, that is, $k_0d = \pi n$ (where $n$ is an integer), the master equation for nonchiral system simplifies to include only the dissipative components $\Gamma_j$ and $\Gamma_{\text{coll},a,b}$. In chiral system, however, in addition to the aforementioned dissipative components, there also exists $g_{a,b}$, representing the interaction strength between atoms, which introduces an additional coherent excitation transfer process between giant atoms.

\begin{table*}
	\centering
	\caption{General expression for $\delta\omega_j$, $g_{a,b}$, $\Gamma_j$, $\Gamma_{\text{coll}, a,b}$ in Eq. (\ref{eq:me}) are denoted as in Ref. \cite{Kockum47, Soro51}. $\phi_{j_n,j_m}(\phi_{a_n,b_m})$ represents the phase shift of two neighbouring coupling points.}
	\begin{tabular}{lll}
		\toprule
		\multicolumn{1}{c}{\textbf{Coefficient}} &   \multicolumn{1}{c}{\textbf{nonchiral coupling}} &  \multicolumn{1}{c}{\textbf{chiral coupling}}
		\\
		\midrule
		\hline
		$\delta\omega_j$ &   $  \sum_{n = 1}^{3} \sum_{m =n =1}^{3} \frac{\sqrt{\gamma_{j_n} \gamma_{j_m}}}{2} \sin \phi_{j_n, j_m}$  & $\sum_{n=1}^{3}\sum_{m=n=1}^{3} \frac{ \left(\sqrt{\gamma_{j_nR}\gamma_{j_mR}} +\sqrt{\gamma_{j_nL}\gamma_{j_mL}}\right)}{2} \sin(\phi_{j_n,j_m})$ \\
		$\Gamma_j$ &  $\sum_{n = 1}^{3} \sum_{m=n=1}^{3} \sqrt{\gamma_{j_n} \gamma_{j_m}} \cos \phi_{j_n, j_m}$ & $\sum_{n=1}^{3} \sum_{m=n=1}^{3} \left( \sqrt{\gamma_{j_nR}\gamma_{j_mR}} + \sqrt{\gamma_{j_nL}\gamma_{j_mL}} \right) \cos(\phi_{j_n,j_m})$ \\
		$\Gamma_{\text{coll}, a,b}$ & $\sum_{n = 1}^{3} \sum_{m=n = 1}^{3} \sqrt{\gamma_{a_n} \gamma_{b_m}} \cos \phi_{a_n, b_m}$ &  $\sum_{n=1}^{3} \sum_{m=n=1}^{3}  \left[ \sqrt{\gamma_{a_nR}\gamma_{b_mR}}\, \exp(\varepsilon i \phi_{a_n,b_m}) + \sqrt{\gamma_{a_nL}\gamma_{b_mL}}\, \exp(- \varepsilon i \phi_{a_n,b_m}) \right]$ \\
		$g_{a,b}$ & $\sum_{n = 1}^{3} \sum_{m =n= 1}^{3} \frac{\sqrt{\gamma_{a_n} \gamma_{b_m}}}{2} \sin \phi_{a_n, b_m}$ & $\sum_{n=1}^{3} \sum_{m=n=1}^{3}  \frac{\varepsilon}{2i} \left[ \sqrt{\gamma_{a_nR}\gamma_{b_mR}}\, \exp(\varepsilon i \phi_{a_n,b_m}) - \sqrt{\gamma_{a_nL}\gamma_{b_mL}}\, \exp(- \varepsilon i \phi_{a_n,b_m}) \right]$ \\
		\bottomrule
	\end{tabular}
	\label{tab:exa}
\end{table*}
General expressions for the parameters in Eq. (\ref{eq:me}) for the nonchiral and chiral cases are listed in Table \ref{tab:exa}. From these expressions, it can be observed that, contrary to the nonchiral scenario, $\Gamma_{\text{coll}, a,b}$ and $g_{a,b}$ in the chiral case are no longer always real; they can also become complex. Additionally, it's also observed that, unlike in small atoms, all parameters for giant atoms depend on the phase shift between their neighbouring coupling points. The sign of $\varepsilon$ is related to the arrangement order of the coupling points between the two giant atoms and the wavuguides.
\begin{equation}
    \varepsilon = \begin{cases} +1 & \text{ if } x_{a_n}<x_{b_m}\\
    \;\;\,0 & \text{ if } x_{a_n}=x_{b_m}\\
    -1 & \text{ if } x_{a_n}>x_{b_m} \end{cases}
    \label{eq:epsilon}
\end{equation}
Clearly, based on the arrangement ordering of the coupling points of two giant atoms with the wavuguide in Fig. \ref{fig1}, and with reference to Table \ref{tab:exa} and Eq. (\ref{eq:epsilon}), we are able to obtain expressions for $\delta\omega_j$, $\Gamma_j$, $g_{a,b}$ and $\Gamma_{\text{coll}, a,b}$ for each configuration in Fig. \ref{fig1}. By choosing specific phase shift values and substituting them into these expressions mentioned above, the corresponding physical mechanism of entanglement generation can be analyzed. It can be understood that the dynamics of entanglement control in a two-qubit system are influenced by two main parameters $g_{a,b}$ and $\Gamma_{\text{coll}, a,b}$ in distinct ways \cite{Gonzalez-Tudela24}. Specifically, $g_{a,b}$ induces oscillations, whereas $\Gamma_{\text{coll}, a,b}$ results in non-oscillatory behavior. Hence, in the discussion that follows, we focus mainly on the effect of these two parameters on entanglement. For convenience, in the following discussion, we denote the interaction strength of the two giant atoms as $g$ and their collective dissipation as $\Gamma_{\text{coll}}$. The Lamb shifts of giant atoms $a$ and $b$ are denoted as $\delta\omega_a$, $\delta\omega_b$, respectively, while the individual decay rates are $\Gamma_a$, and $\Gamma_b$.

Using the Markov quantum master equation for two giant atoms obtained above, we can study the generation of entanglement of two giant atoms. In this paper, we focus only on the case where a giant atom is in an excited state at the initial moment. Consequently, we neglect the jump term in the master equation to derive the non-Hermitian effective Hamiltonian $H_{\text{eff}}$.
\begin{align}
	\label{Heff}
	\hat{H}_{\text{eff}} &= \sum_{j=a,b} \delta \omega_j \sigma^+_j \sigma^-_j + (g\sigma^-_a \sigma^+_b + \text{H.c.})- \frac{i}{2} \sum_{j=a,b} \Gamma_j \sigma^+_j \sigma^-_j - \frac{i}{2} (\Gamma_{\text{coll}} \sigma^-_a \sigma^+_b + \text{H.c.})
\end{align}
Any single-excitation state vector can be expressed as
\begin{equation}
\label{Tstate}
|\psi(t)\rangle = c_{eg}(t) |e\rangle_a |g\rangle_b + c_{ge}(t) |g\rangle_a |e\rangle_b,
\end{equation}
Here, $c_{\text{eg}}(t)$ and $c_{\text{ge}}(t)$ are the probability amplitudes, whose expressions, related to $\delta\omega_j$, $\Gamma_j$, $g$ and $\Gamma_{\text{coll}}$ can be obtained by solving the Schr\"odinger equation with the non-Hermitian effective Hamiltonian $H_{\text{eff}}$. 

For entanglement measurement, various methods exist for quantifying entanglement in two-qubit systems. In this study, we adopt the concurrence method proposed by Wootters \emph{et al.} \cite{Wootters61}. The range of values for concurrence typically falls between $0$ and $1$, where $0$ signifies that the two qubits are in a separable state, and $1$ indicates that the two qubits are in a maximally entangled state. A higher entanglement value corresponds to a concurrence value closer to 1, signifying a stronger correlation between the quantum qubits in the system. In our setup, when we consider two giant atoms initially in the single-excitation state, the concurrence is given by the simple expression
\begin{equation}
C(t) =2|c_{eg}(t) c_{ge}^{\ast }(t)|.\label{Concurrence}
\end{equation}
In the following text, the concurrence of two giant atoms is represented as $C_{eg}^{\text{-nc}}$ ($C_{ge}^{\text{-nc}}$) for nonchiral coupling and $C_{eg}^{\text{-c}}$ ($C_{ge}^{\text{-c}}$) for chiral coupling. It is worth noting that the chirality mentioned here refers to perfect chirality. In the following text, all $\gamma_{kL}$ values are set to be 0 in Table \ref{tab:exa}, except for Sec. \ref{V}.

\section{ENTANGLEMENT GENERATION OF TWO GIANT ATOMS IN BIDIRECTIONAL-CHIRAL WAVEGUIDE}\label{III}
In this section, we explore the spontaneous entanglement generation between two giant atoms within a waveguide-QED system across five different configurations, as illustrated in Fig. \ref{fig1}. We consider giant atom $a$ to be in the excited state and giant atom $b$ in the ground state, denoted as $|e_a,g_b\rangle $. Specifically, to highlight the enhancement of entanglement in chiral case, it is essential to explore the optimal entanglement generation in nonchiral scenarios. Additionally, exploring the influence of the number of coupling points between giant atoms and the waveguide in nonchiral case on entanglement properties is also worthwhile.
\begin{figure*}[!htbp]
	\includegraphics[width=1\textwidth]{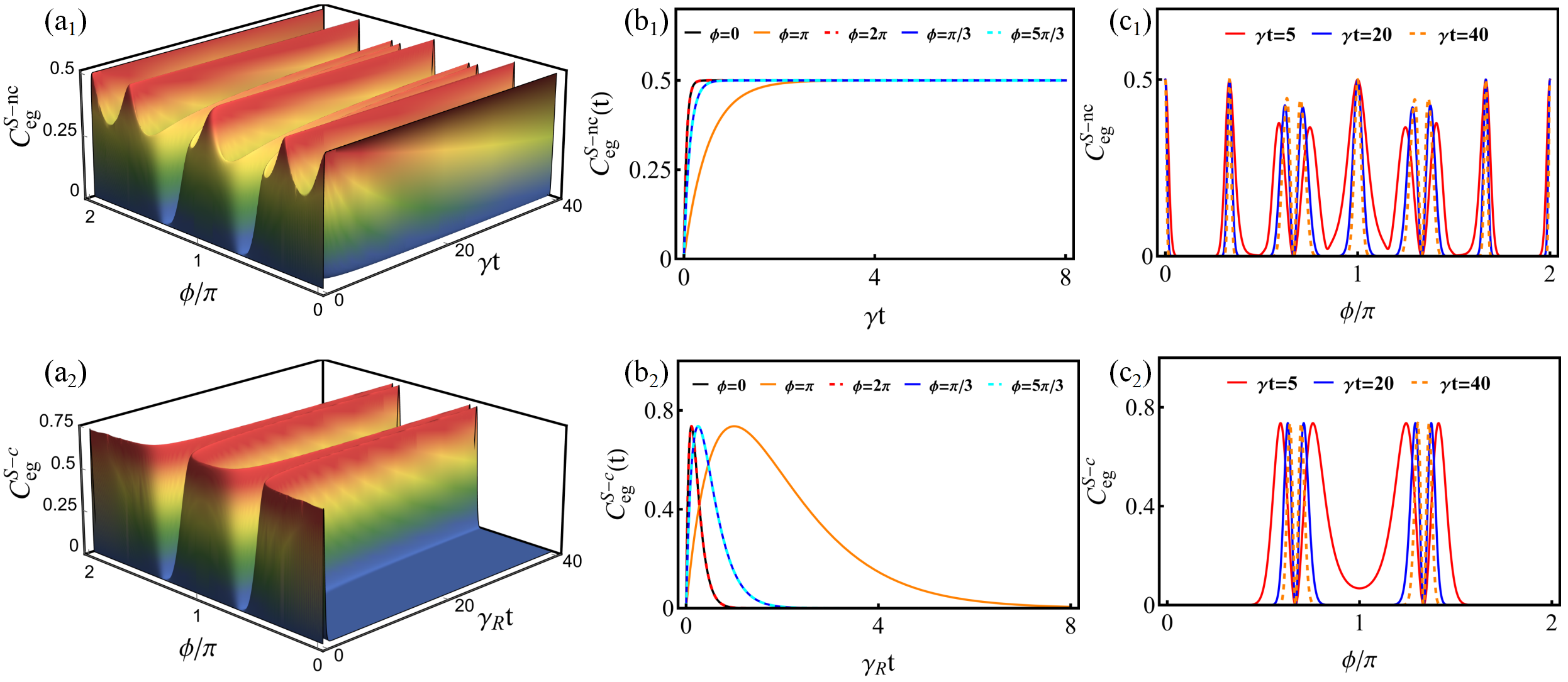}\hfill
	\caption{(Color online)  ($a_{1}$) Concurrence $C_{eg}^{S-nc}$ as functions of $\gamma t$ and $\phi/\pi$, ($a_{2}$) Concurrence $C_{eg}^{S-c}$ as functions of $\gamma_R t$ and $\phi/\pi$. Time evolution of concurrence ($b_{1}$) $C_{eg}^{S-nc}(t)$ and ($b_{2}$) $C_{eg}^{S-c}(t)$ for given phase shift $\phi$. Variation of concurrence ($c_{1}$) $C_{eg}^{S-nc}$ and ($c_{2}$) $C_{eg}^{S-c}$ with phase shift $\phi/\pi$ at specific values of $\gamma t$ and $\gamma_R t$. Curves associated with nonchiral coupling are depicted on the top, and those with chiral coupling are depicted on the bottom, respectively.}
	\label{fig2}
\end{figure*}
\subsection{Entanglement evolution of two giant atoms in separated configuration}\label{Separated configurations}
The separated configuration serves as the first model we introduce, depicted in Fig. \ref{fig1}(a). In Figs. \ref{fig2}($a_{1}$) and \ref{fig2}($a_{2}$), we show the concurrences $C_{eg}^{S-nc}$ and $C_{eg}^{S-c}$ as functions of $\gamma t$ ($\gamma_R t$), and $\phi/\pi$, respectively. From Figs. \ref{fig2}($a_{1}$) and \ref{fig2}($a_{2}$), one can observe that the  concurrences $C_{eg}^{S-nc} $ and $C_{eg}^{S-c}$ exhibit a $2\pi$-periodic dependence on $\phi$. When the phase shift $\phi \in [0, \pi]$, the concurrences $C_{eg}^{S-nc}(t, \phi)$ and $C_{eg}^{S-nc}(t, 2\pi - \phi)$ satisfies the relationship $C_{eg}^{S-nc}(t, \phi) = C_{eg}^{S-nc}(t, 2\pi - \phi)$, and this relation also applies to $C_{eg}^{S-c}$. As shown in Figs. \ref{fig2}($a_{1}$) and \ref{fig2}($a_{2}$), when the phase shift $\phi$ is set to $2\pi/3$ and $4\pi/3$, both concurrences $C_{eg}^{S-nc}$ and $C_{eg}^{S-c} $ are zero. From the preceding discussion and the information provided in Table \ref{tab:exa}, it is evident that the interaction strength $g$, individual decay rates $\Gamma_a$ and $\Gamma_b$, as well as collective dissipation $\Gamma_{\text{coll}}$, are all influenced by the phase shift $\phi$. Notably, for the phase shift $\phi$ taken as $2\pi/3$ and $4\pi/3$, all the parameters mentioned above are zero, i.e., the excitation paths of the two giant atoms interfere destructively at these two phase shifts. Consequently, two giant atoms are entirely decoupled from the waveguide, resulting in the absence of interatomic entanglement.

To better understand how phase shifts modulate entanglement and the influence of chiral effects on entanglement property, we plot the time evolution of concurrences $C_{eg}^{S-nc}(t)$ and $C_{eg}^{S-c}(t)$ for specific values of $\phi$ in the range $\phi\in [0, 2\pi]$, respectively. When the phase shift $\phi$ is set to $n\pi$ (for an integer $n$), $\pi/3$, or $5\pi/3$, as shown in Fig. \ref{fig2}($b_{1}$), the value of concurrence $C_{eg}^{S-nc}(t)$ is always maintained at 0.5, indicating that steady-state entanglement can be achieved. The reason for this is, when phase shift $\phi$ = $n\pi$, $\pi/3$ and $5\pi/3$, interaction strength $g\rightarrow 0$ of induced oscillatory behaviour, steady-state behavior of the concurrence $C_{eg}^{S-nc}(t)$ is predominantly influenced by collective dissipation $\Gamma_{\text{coll}}$. The appearance of steady-state entanglement is mainly due to the presence of long-lived dark state. It should be noted that in this configuration, due to quantum interference effects among multiple coupling points, increasing the number of coupling points allows us to achieve steady-state entanglement under a greater number of phase shifts, in comparison with Ref \cite{Yin55}.
Fig. \ref{fig2}($b_{2}$) depicts the entanglement dynamics for the giant atoms chirally coupled to the waveguide. A comparison with Fig. \ref{fig2}($b_{1}$) reveals differences in the behaviors of concurrences $C_{eg}^{S-c}(t)$ and $C_{eg}^{S-nc}(t)$. In the case of chiral coupling, there is no steady-state entanglement when $\phi = n\pi$, $\pi/3$, and $5\pi/3$. The concurrence $C_{eg}^{S-c}(t)$ quickly peaks at 0.736 before rapidly decreasing to 0. Indeed, the chiral coupling modifies the entanglement generation mechanism, which, in the chiral case, no longer relies solely on the incoherent part $\Gamma_{\text{coll}}$. It includes an additional term, $g$, that facilitates a coherent transfer of excitations between the atoms. This extra $g$ is responsible for the greater concurrence observed under chiral coupling in this configuration. 

In order to better observe the joint influence of the quantum interference and chiral effects on entanglement dynamics, the concurrences $C_{eg}^{S-nc}$ and $C_{eg}^{S-c}$ are plotted against $\phi/\pi$ for specific $\gamma t$ and $\gamma_R t$ values, where $\phi\in [0, 2\pi]$. In Fig. \ref{fig2}($c_{1}$), the peaks of concurrence $C_{eg}^{S-nc}$ predominantly manifest at $\phi =n\pi$, $\pi/3$, $5\pi/3$. Furthermore, when $\gamma t$ takes larger values, the concurrence $C_{eg}^{S-nc}$ demonstrates a gradual increase when $\phi$ is close to $2\pi/3$ and $4\pi/3$. In contrast, as illustrated in Fig. \ref{fig2}($c_{2}$), when $\phi$ approaches $2\pi/3$ and $4\pi/3$, the value of concurrence $C_{eg}^{S-c}$ remains at 0.736 as $\gamma_R t$ increases. 
\subsection{Entanglement evolution of two giant atoms in fully braided configuration}\label{Fully braided configurations}
\begin{figure*}[!htbp]
	\includegraphics[width=1\textwidth]{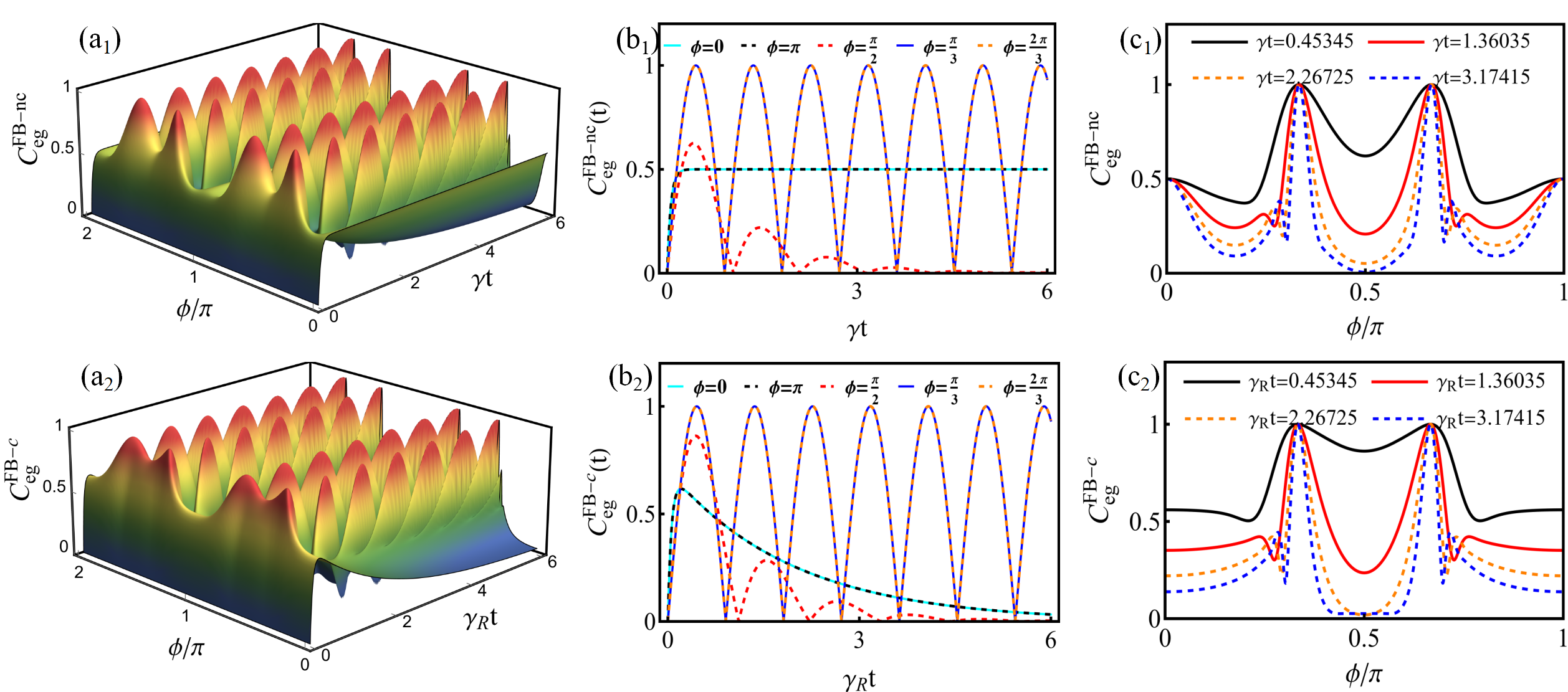}\hfill
	\caption{(Color online)  ($a_{1}$) Concurrence $C_{eg}^{FB-nc}$ as functions of $\gamma t$ and $\phi/\pi$, ($a_{2}$) Concurrence $C_{eg}^{FB-c}$ as functions of $\gamma_R t$ and $\phi/\pi$. Time evolution of concurrence ($b_{1}$) $C_{eg}^{FB-nc}(t)$ and ($b_{2}$) $C_{eg}^{FB-c}(t)$ for given phase shift $\phi$. Variation of concurrence ($c_{1}$) $C_{eg}^{FB-nc}$ and ($c_{2}$) $C_{eg}^{FB-c}$ with phase shift $\phi/\pi$ at specific values of $\gamma t$ and $\gamma_R t$. Curves associated with nonchiral coupling are depicted on the top, and those with chiral coupling are depicted on the bottom, respectively.}
	\label{fig3}
\end{figure*}

In this subsection, we introduce the fully braided configuration, as depicted in Fig. \ref{fig1}(b). Similar to the case of two separate giant atoms, we examine the entanglement generated between two braided atoms under both nonchiral (top) and chiral (bottom) coupling case. In Figs. \ref{fig3}($a_{1}$) and \ref{fig3}($a_{2}$), we show the evolution of concurrences $C_{\text{eg}}^{FB-nc}$ and $C_{\text{eg}}^{FB-c}$, respectively, as functions of $\gamma t$ ($\gamma_R t$) and $\phi/\pi$. Different from the separated configuration, the concurrences $C_{\text{eg}}^{FB-nc}$ and $C_{\text{eg}}^{FB-c}$ exhibit phase dependence with a period of $\pi$.

We know that $g$ and $\Gamma_{\text{coll}}$ significantly affect entanglement dynamics. In the following, we will discuss how the two parameters influence it. To do this, we plot the dynamics evolution of concurrences $C_{\text{eg}}^{FB-nc}(t)$ and $C_{\text{eg}}^{FB-c}(t)$ when $\phi$ takes some specific values. As shown in Fig. \ref{fig3}($b_{1}$), the concurrence $C_{\text{eg}}^{FB-nc}(t)$ exhibits the characteristic of approaching a steady-state value 0.5 at $\phi = 0$, $\pi$. In this scenario, the oscillations produced by $g$ are completely suppressed ($g = 0$); collective dissipation $\Gamma_{\text{coll}}$, which induces non-oscillatory behaviour, plays a dominant role ($\Gamma_{\text{coll}}\neq 0$). Moreover, similar to the separated configuration, the emergence of steady-state entanglement predominantly stems from the existence of long-lived dark state. For phase shift $\phi = \pi/2$, the concurrence $C_{\text{eg}}^{FB-nc}(t)$ primarily exhibits oscillatory decay features. When $\phi$ is set to $\pi/3$, $2\pi/3$, the concurrence $C_{\text{eg}}^{FB-nc}(t)$ exhibits a behavior of initially rising to its maximum value 1, followed by a rapid drop to 0, periodically oscillating between 0 and 1, as shown in Fig. \ref{fig3}($b_{1}$). This is because when $\phi = \pi/$3 and $2\pi/3$, $\Gamma_a$ = $\Gamma_b$ = $\Gamma_{\text{coll}}$ = $0$, meaning no photons are emitted from the two-giant atom system into the waveguide, but, the atomic interaction strength $g$ (referred to as the decoherence-free (DF) interaction \cite{Kockum47}) of the induced oscillations still exists. In this case, the concurrence $C_{\text{eg}}^{FB-nc}(t)$ exhibits primarily oscillatory behavior, and its maximum entanglement can reach 1. Compared with Ref. \cite{Yin55}, we have one more phase shift that can achieve the maximum entanglement value 1. In the case of chirality, as one can see from Fig. \ref{fig3}($b_{2}$), when $\phi = 0$, $\pi$, the value of  concurrence $C_{\text{eg}}^{FB-c}(t)$ no longer remains at 0.5. Instead, it exhibits a unique behavior in that it initially rises to values greater than 0.5 and then rapidly falls to 0, which differs from the features of concurrence $C_{\text{eg}}^{FB-nc}(t)$ in the nonchiral case. When we substitute these two phase shifts into the general expression of $g$ and $\Gamma_{\text{coll}}$ under chiral coupling, we find that both $g$ and $\Gamma_{\text{coll}}$ are non-zero. Hence, the concurrence $C_{\text{eg}}^{FB-c}(t)$ experiences a rapid increase due to the nonzero interaction strength $g$, followed by a swift decay caused by the nonzero collective dissipation $\Gamma_{\text{coll}}$ of two giant atoms.
In particular, since chirality has no effect on the DF interaction \cite{Carollo, Soro51}, we have made a natural discovery: at $\phi= \pi/3$ and $2\pi/3$, the behavior of concurrence $C_{\text{eg}}^{FB-c}(t)$ is consistent with that of $C_{\text{eg}}^{FB-nc}(t)$. That is, the entanglement maximum value in the chiral case can also reach 1 when $\phi= \pi/3$ and $2\pi/3$.

Figures \ref{fig3}($c_{1}$) and \ref{fig3}($c_{2}$) depict concurrences $C_{\text{eg}}^{FB-nc}$ and $C_{\text{eg}}^{FB-c}$ as a function of $\phi/\pi$ for specific values of $\gamma t$ and $\gamma_R t $. By comparing Figs. \ref{fig3}($c_{1}$) and \ref{fig3}($c_{2}$), one can clearly see that concurrences $C_{\text{eg}}^{FB-nc}$ and $C_{\text{eg}}^{FB-c}$ share some common behaviors in their overall trends. Within the phase shift range of $\phi \in [0, \pi]$, entanglement peak values for both concurrences $C_{\text{eg}}^{FB-nc}$ and $C_{\text{eg}}^{FB-c}$ are observed at $\phi = \pi/3$ and $2\pi/3$. Furthermore, it is clear that the width of the peaks in concurrences $C_{\text{eg}}^{FB-nc}$ and $C_{\text{eg}}^{FB-c}$ decreases as $\gamma t$ and $\gamma_R t$ increase. The difference is, the concurrence $C_{\text{eg}}^{FB-nc}$ remains at 0.5 for $\phi = 0$ and $\pi$, indicating its insensitivity to variations in $\gamma t$, as depicted in Fig. \ref{fig3}($c_{1}$). This contrasts with the behavior of concurrence $C_{\text{eg}}^{FB-c}$ in Fig. \ref{fig3}($c_{2}$), where its entanglement value decreases with increasing $\gamma_R t$ at $\phi = 0$ and $\pi$.
\begin{figure*}[!htbp]
	\includegraphics[width=1\textwidth]{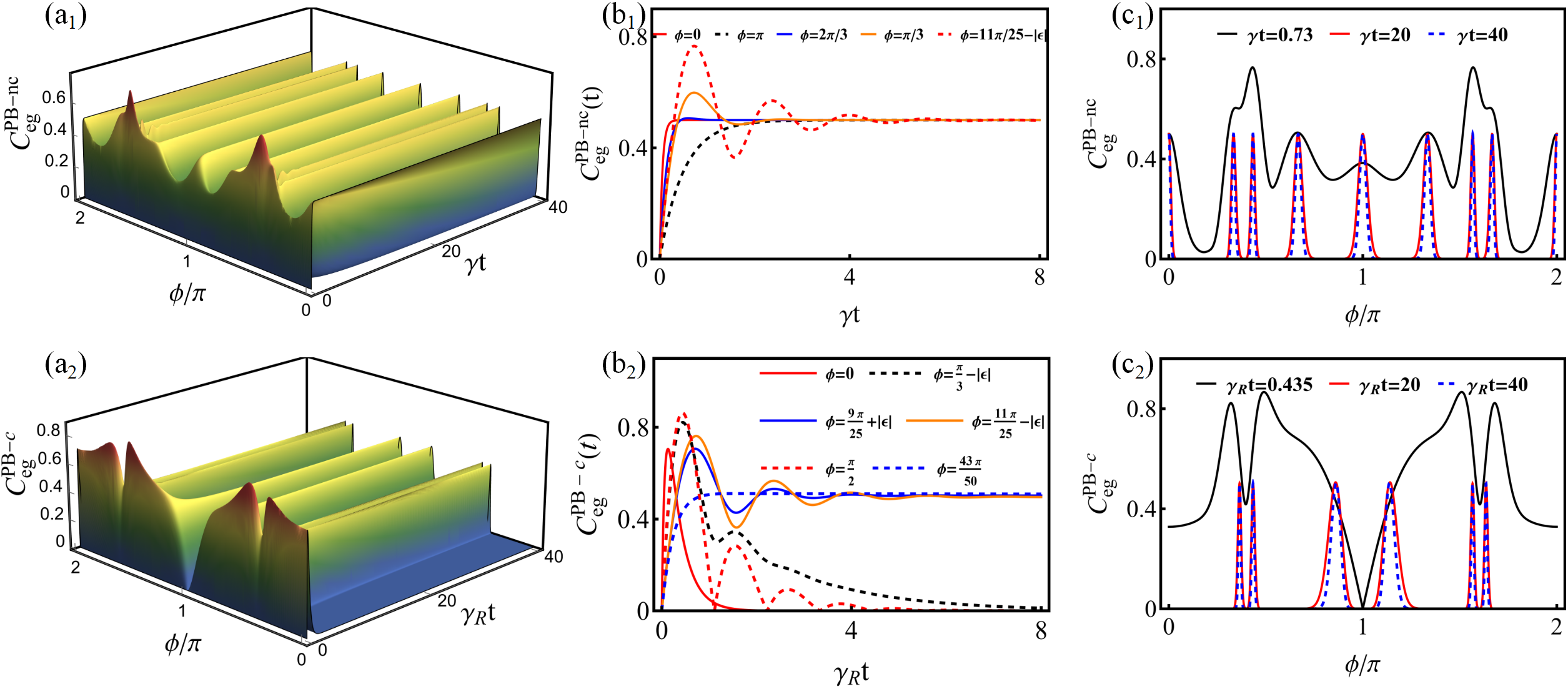}\hfill
	\caption{(Color online)  ($a_{1}$) Concurrence $C_{eg}^{PB-nc}$ as functions of $\gamma t$ and $\phi/\pi$, ($a_{2}$) Concurrence $C_{eg}^{PB-c}$ as functions of $\gamma_R t$ and $\phi/\pi$. Time evolution of concurrence ($b_{1}$) $C_{eg}^{PB-nc}(t)$ and ($b_{2}$) $C_{eg}^{PB-c}(t)$ for given phase shift $\phi$. Variation of concurrence ($c_{1}$) $C_{eg}^{PB-nc}$ and ($c_{2}$) $C_{eg}^{PB-c}$ with phase shift $\phi/\pi$ at specific values of $\gamma t$ and $\gamma_R t$. Curves associated with nonchiral coupling are depicted on the top, and those with chiral coupling are depicted on the bottom, respectively.}
	\label{fig4}
\end{figure*}
\subsection{Entanglement evolution of two giant atoms in partially braided configuration}\label{Partially braided configurations}

Now, let us introduce another braided configuration: partially braided, as shown in Fig. \ref{fig1}(c). We depict the concurrences $C_{\text{eg}}^{PB-nc}$ and $C_{\text{eg}}^{PB-c}$ as functions of $\gamma t$ ($\gamma_R t$) and $\phi/\pi$ in Figs. \ref{fig4}($a_{1}$) and \ref{fig4}($a_{2}$). It can be observed that the entanglement dynamics of concurrence $C_{\text{eg}}^{PB-nc}$ ($C_{\text{eg}}^{PB-c}$) exhibit mirror symmetry between the ranges $\phi\in [0, \pi]$ and $\phi\in [\pi, 2\pi]$. Different from the fully braided configuration, the concurrences $C_{\text{eg}}^{PB-nc}$ and $C_{\text{eg}}^{PB-c}$ exhibit a $2\pi$-periodic dependence on $\phi$, as demonstrated in Figs. \ref{fig4}($a_{1}$) and \ref{fig4}($a_{2}$). Additionally, we can also find that both concurrences $C_{\text{eg}}^{PB-nc}$ and $C_{\text{eg}}^{PB-c}$ tend to reach a steady value 0.5 at certain phase shifts within the consider timescale, which also markedly differs from the fully braided configuration.

\begin{figure*}[!htbp] 
\includegraphics[width=1\textwidth]{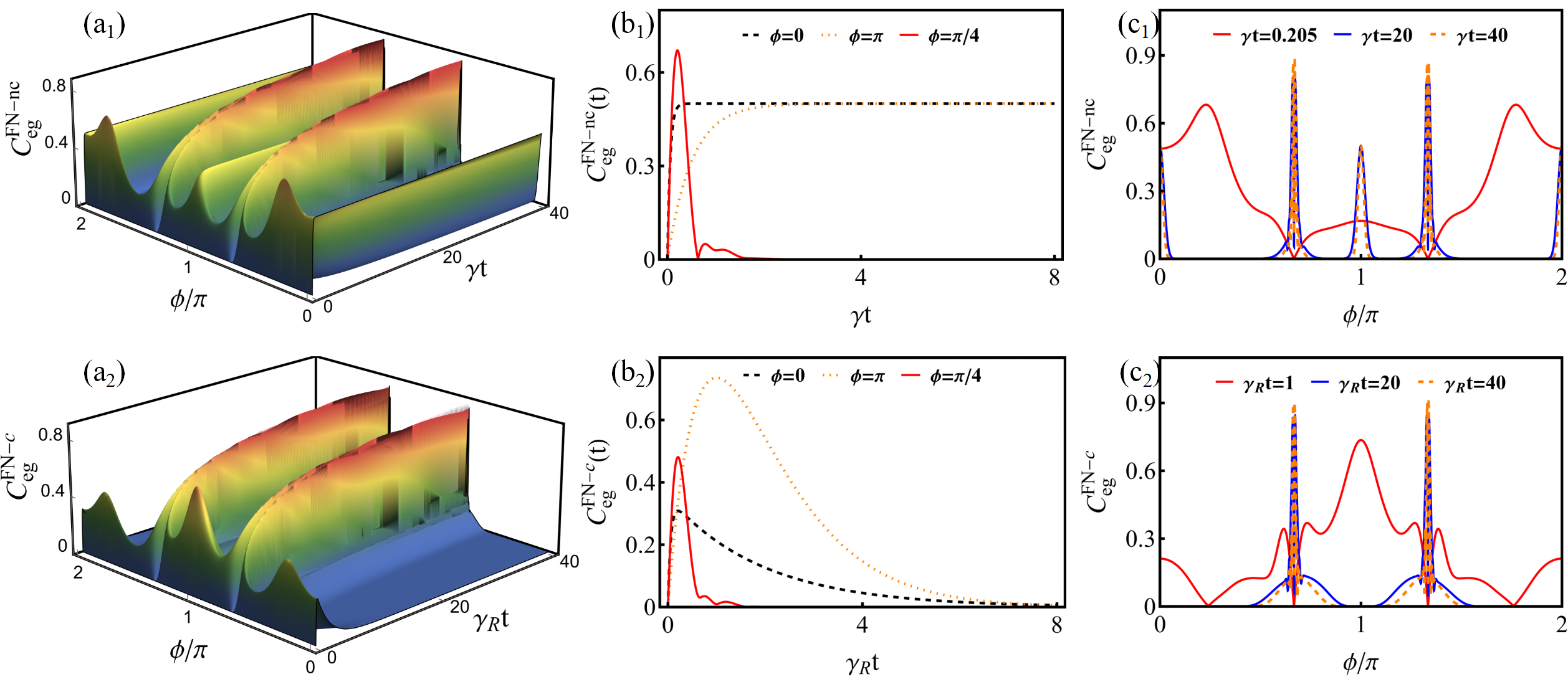}\hfill
\caption{(Color online)  ($a_{1}$) Concurrence $C_{eg}^{FN-nc}$ as functions of $\gamma t$ and $\phi/\pi$, ($a_{2}$) Concurrence $C_{eg}^{FN-c}$ as functions of $\gamma_R t$ and $\phi/\pi$. Time evolution of concurrence ($b_{1}$) $C_{eg}^{FN-nc}(t)$ and ($b_{2}$) $C_{eg}^{FN-c}(t)$ for given phase shift $\phi$. Variation of concurrence ($c_{1}$) $C_{eg}^{FN-nc}$ and ($c_{2}$) $C_{eg}^{FN-c}$ with phase shift $\phi/\pi$ at specific values of $\gamma t$ and $\gamma_R t$. Curves associated with nonchiral coupling are depicted on the top, and those with chiral coupling are depicted on the bottom, respectively.}
\label{fig5}
\end{figure*}

In order to better compare the influence of quantum interference effects on the entanglement dynamics in two coupling cases, we show the time evolution of concurrences $C_{\text{eg}}^{PB-nc}(t)$ and $C_{\text{eg}}^{PB-c}(t)$ as $\phi$ takes different values, where $\phi\in [0, \pi]$. In the nonchiral case, as can be seen directly in Fig. \ref{fig4}($b_{1}$), when $\phi = 0, \pi$, concurrence $C_{\text{eg}}^{PB-nc}(t)$ approaches steady value 0.5, and the entangled state primarily originates from the dark state. The physical mechanism remains in line with the ones in both separated and fully braided configurations. In addition, due to the complex quantum interference effects among multiple coupling points, when $\phi$ takes certain values other than integer multiples of $\pi$, such as $\phi=\pi/3$, $2\pi/3$, the value of concurrence $C_{\text{eg}}^{PB-nc}(t)$ is also fixed at $0.5$, as demonstrated in Fig. \ref{fig4}($b_{1}$), a phenomenon not observed in other configurations mentioned in this paper.
Specifically, the maximal value of concurrence $C_{\text{eg}}^{PB-nc}(t)$ is limited to 0.77 at $\phi = 11\pi/25 - \left|\epsilon\right|$, with $\epsilon \ll 1$ (a same define will be adopted for subsequent text), as depicted in Fig. \ref{fig4}($b_{1}$). At this phase shift, the interaction strength $g$ leads to a rapid increase in quantum entanglement, reaching a peak value 0.77 in a short time. Nonetheless, a slight oscillatory decay behavior follows because of the presence of the collective dissipation $\Gamma_{\text{coll}}$, and the entanglement ultimately converges to the steady-state value 0.5 as a result of the combined effect of these two parameters. In the case of chiral coupling, as illustrated in Fig. \ref{fig4}($b_{2}$), compared to nonchiral coupling, an additional $g$ serves as a hindrance to the formation of the dark state for phase shifts $\phi = 0$, therefore, there is no steady-state entanglement generation at this phase shift, as explained in the separated and fully braided configurations. However, at some other phase shifts, such as $\phi=11\pi/25 - \left|\epsilon\right|$, due to the collective effects of chirality and quantum interference, the concurrence $C_{\text{eg}}^{PB-c}(t)$ remains at steady-state value 0.5. This feature is distinct from other configurations and small atom when chirality is considered. Although steady-state entanglement is observed in both Figs. \ref{fig4}($b_{1}$) and \ref{fig4}($b_{2}$), a direct comparison reveals that phase shifts associated with steady-state entanglement in chiral coupling, where $\phi \neq n\pi$, do not correspond with those in nonchiral coupling. For example, phase shifts like $9\pi/25 +\left|\epsilon\right|$, $43\pi/50$ and $\pi/3-\left|\epsilon\right|$, which contribute to the generation of steady-state entanglement in the chiral coupling shown in Fig. \ref{fig4}($b_{2}$), do not yield such entanglement in the nonchiral case of Fig. \ref{fig4}($b_{1}$). As well, due to the presence of chirality, in Fig. \ref{fig4}($b_{2}$) with the phase shift $\phi = \pi/2$, the concurrence $C_{\text{eg}}^{PB-c}(t)$ enables a maximum entanglement 0.87, which is higher than the 0.77 achieved in the nonchiral case.

For a clearer observation of the phase shift $\phi$ modulation on concurrences $C_{\text{eg}}^{PB-nc}$ and $C_{\text{eg}}^{PB-c}$, we present these concurrences as functions of $\phi$ for various $\gamma t$ and $\gamma_R t $ values in Figs. \ref{fig4}($c_{1}$) and \ref{fig4}($c_{2}$).
In the nonchiral coupling case, concurrence $C_{\text{eg}}^{PB-nc}$ initially exhibits two peaks at early times (e.g., $\gamma t = 0.73$), which peak values correspond to the entanglement maximum value. As time extends (e.g., $\gamma t = 20$), the concurrence $C_{\text{eg}}^{PB-nc}$ stabilizes at 0.5 for several phase shifts, and the previously mentioned peaks disappear. With further increase in $\gamma t$, the blue dotted and the red solid lines overlap in Fig. \ref{fig4}($c_{1}$), the concurrence $C_{\text{eg}}^{PB-nc}$ consistently remains at 0.5, showing that entangled states have long lifetimes. Moreover, it is clear that when $\gamma t$ takes a large value, there is no entanglement between the two giant atoms at other phase shifts, except for steady state entanglement at specific phase shifts. 
For chiral coupling, as illustrated in Fig. \ref{fig4}($c_{2}$), when $\gamma_R t$ is small (e.g., $\gamma_R t = 0.435$), we observe two main peaks along with two sub-peaks, all of which have peak values larger than those observed in the nonchiral coupling case. In addition, at the phase shift $\phi= \pi$, the concurrence $C_{\text{eg}}^{PB-c}$ is always 0, a behavior that is independent of $\gamma_R t$ and differs from the one observed in Fig. \ref{fig4}($c_{1}$), where concurrence $C_{\text{eg}}^{PB-nc}$ exhibits steady-state properties. However, for concurrence $C_{\text{eg}}^{PB-c}$ in Fig. \ref{fig4}($c_{2}$), steady entanglement can be obtained when the phase shift $\phi$ approaches $\pi$ from both sides.

\begin{figure*}[!htbp]
	\includegraphics[width=1\textwidth]{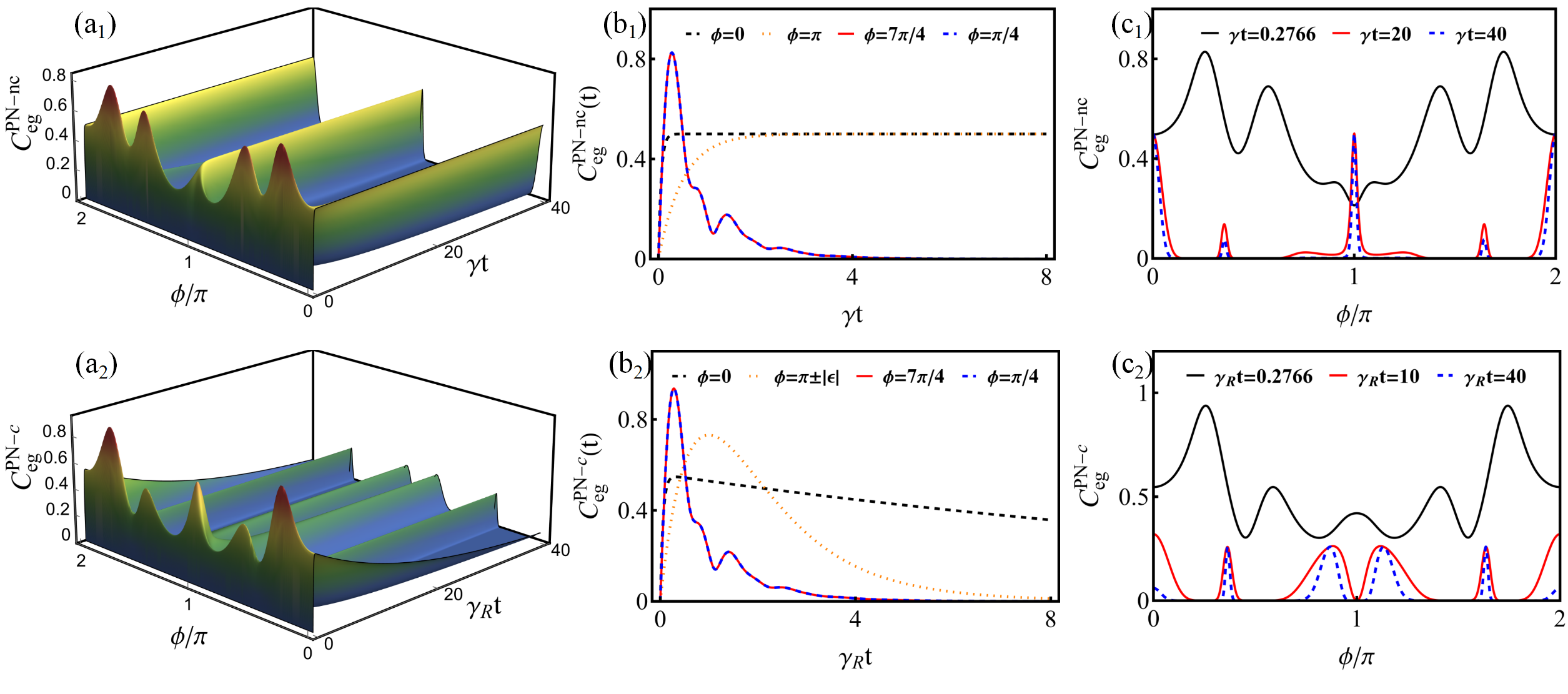}\hfill
	\caption{(Color online)  ($a_{1}$) Concurrence $C_{eg}^{PN-nc}$ as functions of $\gamma t$ and $\phi/\pi$, ($a_{2}$) Concurrence $C_{eg}^{PN-c}$ as functions of $\gamma_R t$ and $\phi/\pi$. Time evolution of concurrence ($b_{1}$) $C_{eg}^{PN-nc}(t)$ and ($b_{2}$) $C_{eg}^{PN-c}(t)$ for given phase shift $\phi$. Variation of concurrence ($c_{1}$) $C_{eg}^{PN-nc}$ and ($c_{2}$) $C_{eg}^{PN-c}$ with phase shift $\phi/\pi$ at specific values of $\gamma t$ and $\gamma_R t$. Curves associated with nonchiral coupling are depicted on the top, and those with chiral coupling are depicted on the bottom, respectively.}
	\label{fig6}
\end{figure*}
\subsection{Entanglement evolution of two giant atoms in fully nested configuration}\label{Fully nested configurations}
Following our discussion on entanglement generation in separated and two braided configurations, the focus shifts to generation of entanglement in nested configurations.
For fully nested configuration in Fig. \ref{fig1}(d), the concurrences $C_{\text{eg}}^{FN-nc}$ and $C_{\text{eg}}^{FN-c}$ are present as functions of $\gamma t$ ($\gamma_R t $) and $\phi/\pi$ in Figs. \ref{fig5}($a_{1}$) and \ref{fig5}($a_{2}$). Similar to the separated and partially braided configurations, the concurrences $C_{\text{eg}}^{FN-nc}$ and $C_{\text{eg}}^{FN-c}$ are also phase dependent with a period 2$\pi$ and satisfy the relation $C_{\text{eg}}^{FN-nc} (C_{\text{eg}}^{FN-c}) (t, \phi)$ = $C_{\text{eg}}^{FN-nc} (C_{\text{eg}}^{FN-c}) (t, 2\pi-\phi)$, where $\phi\in [0, \pi]$. It is found that both concurrences $C_{\text{eg}}^{FN-nc}$ and $C_{\text{eg}}^{FN-c}$ can be generated at $\phi= n\pi$ for an integer $n$. In addition, four ridges are observed in Figs. \ref{fig5}($a_{1}$) and \ref{fig5}($a_{2}$) when $\phi$ nears $2\pi/3$ and $4\pi/3$, where both concurrences $C_{\text{eg}}^{FN-nc}$ and $C_{\text{eg}}^{FN-c}$ demonstrate a gradual increase as time evolves. This differs from the two nested giant atoms coupled with the bidirectional waveguide at two points each, which exhibits only two ridge at $\phi \rightarrow \pi$ \cite{Yin55}. 

To clearly observe these features, we present the time evolution of the concurrences $C_{eg}^{FN-nc}(t)$ and $C_{eg}^{FN-c}(t)$ for specific values of the phase shift $\phi$ within the region of $\phi\in [0, \pi]$ in Figs. \ref{fig5}($b_{1}$) and \ref{fig5}($b_{2}$).
Specifically, we observe that the concurrence $C_{\text{eg}}^{FN-nc}(t)$ reaches steady-state value 0.5 at $\phi = 0$ and $\pi$, as depicted in Fig. \ref{fig5}($b_{1}$). This behavior is analogous to what we observed those in the separated and two braided configurations under nonchiral conditions. Moreover, Fig. \ref{fig5}($a_{1}$) reveals that the value of the generated atomic entanglement exceed 0.5 at certain phase shift $\phi$. For instance, in Fig. \ref{fig5}($b_{1}$), it can be seen that the concurrence rapidly reaches a peak value, $C_{\text{eg}}^{FN-nc}(t) \approx 0.67$ at $\phi = \pi/4$, and then undergoes a swift decay to 0. To explain this feature, we set $\phi = \pi/4$ and observe that both $g$ and $\Gamma_{\text{coll}}$ become non-zero. In this scenario, concurrence $C_{\text{eg}}^{FN-nc}(t)$ demonstrates a rapid increase (attributed to the non-zero interaction strength $g$), succeeded by a swift decay (owing to the non-zero collective dissipation $\Gamma_{\text{coll}}$). However, compared with $C_{\text{eg}}^{FN-nc}(t)$, the concurrence $C_{\text{eg}}^{FN-c}(t)$ actually decreases in Fig. \ref{fig5}($b_{2}$) when $\phi = \pi/4$, since the chirality effects are present. Moreover, at $\phi = 0$, concurrence $C_{\text{eg}}^{FN-c}(t)$ falls below $C_{\text{eg}}^{FN-nc}(t)$ in Fig. \ref{fig5}($b_{1}$), whereas surpasses it at $\phi = \pi$. This suggests that in the configuration, the introduction of $g$ is not always beneficial for enhancing entanglement.

Figures. \ref{fig5}($c_{1}$) and \ref{fig5}($c_{2}$) show the concurrences $C_{\text{eg}}^{FN-nc}$ and $C_{\text{eg}}^{FN-c}$, respectively, as functions of $\phi/\pi$ at given values of $\gamma t$ and $\gamma_R t $, where $\phi\in [0, 2\pi]$. In the case of nonchiral coupling, we find that the concurrence $C_{\text{eg}}^{FN-nc}$ can maintain its peak value 0.5 for larger values of $\gamma t$ when $\phi = 0$, $\pi$ and $2\pi$. In addition, concurrence $C_{\text{eg}}^{FN-nc}$ reaches its peak value 0.87 more slowly when $\phi \rightarrow 2\pi/3$ and $4\pi/3$, as shown in Fig. \ref{fig5}($c_{1}$). In the case of chiral coupling, there is one peak in $C_{\text{eg}}^{FN-c}$ within one period when $\gamma_R t $ = 1. For larger values of $\gamma_R t$, e.g., $\gamma_R t =40$, the maximum values of concurrence $C_{\text{eg}}^{FN-c}$ = 0.9 is also mainly created at $\phi \rightarrow 2\pi/3$ and $4\pi/3$. Through comparing the maximum entanglement values between chiral and nonchiral coupling scenarios in this configuration, it is evident that the maximum concurrence achieved through chiral coupling does not show a substantial improvement over nonchiral coupling. Chiral effects offer little apparent advantage for enhancing entanglement in this configuration.

\subsection{Entanglement evolution of two giant atoms in partially nested configuration}\label{Partially nested configurations}
The final configuration under consideration in this article is the partial nested one, as shown in Fig. \ref{fig1}(e). We plot the evolution of concurrences $C_{\text{eg}}^{PN-nc}$ and $C_{\text{eg}}^{PN-c}$ as functions of $\gamma t$ ($\gamma_R t$) and $\phi /\pi$ in Figs. \ref{fig6}($a_{1}$) and  \ref{fig6}($a_{2}$) with the aim to better observe the impact of chiral effects on entanglement. The evolution of concurrences $C_{\text{eg}}^{PN-nc}$ and $C_{\text{eg}}^{PN-c}$ is $2\pi$-periodic in $\phi$ and satisfy the relation $C_{\text{eg}}^{PN-nc} (C_{\text{eg}}^{PN-c}) (t, \phi)$ = $C_{\text{eg}}^{PN-nc} (C_{\text{eg}}^{PN-c}) (t, 2\pi-\phi)$ for $\phi\in [0, \pi]$. From Figs. \ref{fig6}($a_{1}$) and \ref{fig6}($a_{2}$), it can be seen that, similar to the separated, two braided and fully nested configurations, at phase shifts $\phi = n\pi$ (where $n$ is an integer), the concurrence $C_{\text{eg}}^{PN-nc}$ approximates a steady-state value 0.5; whereas the concurrence $C_{\text{eg}}^{PN-c}$ decays to 0 over time, exhibiting transient entanglement characteristics. Additionally, within one period, the concurrence $C_{\text{eg}}^{PN-nc}$ exhibits four peak values greater than 0.5, while concurrence $C_{\text{eg}}^{PN-c}$ has only three. 

we present the dynamic evolution of concurrences $C_{\text{eg}}^{PN-nc}(t)$ and $C_{\text{eg}}^{PN-c}(t)$ at specific phase shifts $\phi$ within the range of [0, 2$\pi$] in Figs. \ref{fig6}($b_{1}$) and \ref{fig6}($b_{2}$).
In the case of nonchiral coupling, as shown in Fig. \ref{fig6}($b_{1}$), when $\phi = 0,\pi$, the concurrence $C_{\text{eg}}^{PN-nc}(t)$ converging to a steady-state value 0.5. The physical mechanism underlying the generation of steady-state entanglement has been discussed in detail in the above configurations, and we will not repeat further here. When the phase shift $\phi$ is set at $\pi/4$ and $7\pi/4$, the concurrence $C_{\text{eg}}^{PN-nc}(t)$ rapidly rises to its peak value 0.83, driven by the non-zero interaction strength $g$. Subsequently, it undergoes oscillations decay to 0 within a short time due to the influence of non-zero collective dissipation $\Gamma_{\text{coll}}$.
Under chiral coupling condition, for the phase shift of $\phi = \pi/4$ and $7\pi/4$, in Fig. \ref{fig6}($b_{2}$), the concurrence $C_{\text{eg}}^{PN-c}(t)$ also exhibits characteristics of oscillatory decay, whereas its maximum entanglement value can reach 0.94. As we mentioned in the separated and fully braided configurations, the coherent excitation transfer introduced by $g$ in the chiral case also induces greater interatomic entanglement in this configuration. In such case, the values of concurrence $C_{\text{eg}}^{PN-c}(t)$ are 0.55 and 0.73 when the phase shift is taken as $\phi=0$ and $\pi+\left|\epsilon\right|$, respectively. Notably, by comparing the entanglement maxima for the two coupling cases in this configuration, it is found that the chiral effect plays a more substantial role in enhancing entanglement compared to the fully nested configuration.

For the purpose of seeing the influence of chiral effects on entanglement properties, we show in Figs. \ref{fig6}($c_{1}$) and \ref{fig6}($c_{2}$) the concurrences $C_{\text{eg}}^{PN-nc}$ and $C_{\text{eg}}^{PN-c}$ as a function of $\phi/\pi$ when $\gamma t$ and $\gamma_R t$ take some particular values. In the nonchiral case depicted in Fig. \ref{fig6}($c_{1}$), the concurrence $C_{\text{eg}}^{PN-nc}$ exhibits a clear minimum value below 0.5 at $\phi = \pi$ for a small $\gamma t$ value of $0.2766$, and as $\gamma t$ increases, the concurrence converges to 0.5. However, when $\phi$ is taken as $0$ and $2\pi$, the concurrence $C_{\text{eg}}^{PN-nc}$ always maintains the steady value 0.5, insensitive to $\gamma t$. Furthermore, it is notable that for $\phi \neq n\pi$, the values of concurrence $C_{\text{eg}}^{PN-nc}$ are relatively small, approaching 0 over a broad range of phase shifts as $\gamma t$ increases.
When two partially nested giant atoms are chirally coupled to the waveguide, as depicted in Fig. \ref{fig6}($c_{2}$), the concurrence $C_{\text{eg}}^{PN-c}$ can achieve maximum values at two phase shifts $\phi$ =$\pi/4$ and $7\pi/4$. Specifically, when phase shifts $\phi=0$ and $\pi$, at the initial moment, the concurrence $C_{\text{eg}}^{PN-c}$ go beyond 0.5. However, as $\gamma_R t$ increases, the concurrence $C_{\text{eg}}^{PN-c}$ falls below 0.5 over the entire period.

\section{EFFECT OF INITIAL STATE ON THE DYNAMICS OF ENTANGLEMENT OF TWO GIANT ATOMS}\label{IV}
\begin{figure}[tbp]
	\centering
	\includegraphics[width=0.48\textwidth]{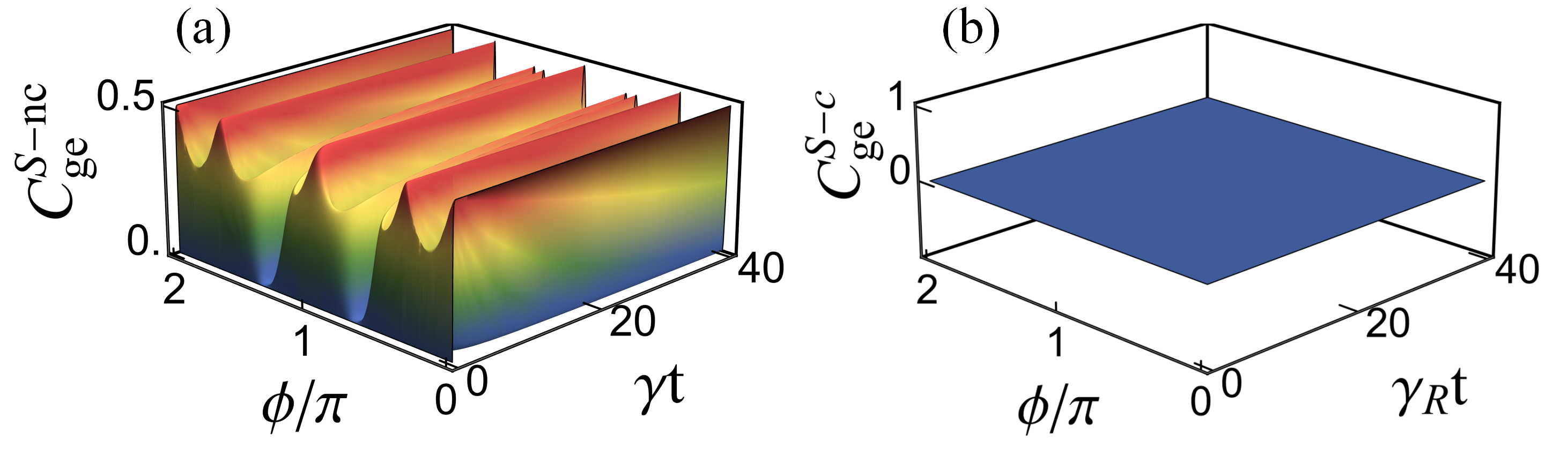}
	\includegraphics[width=0.48\textwidth]{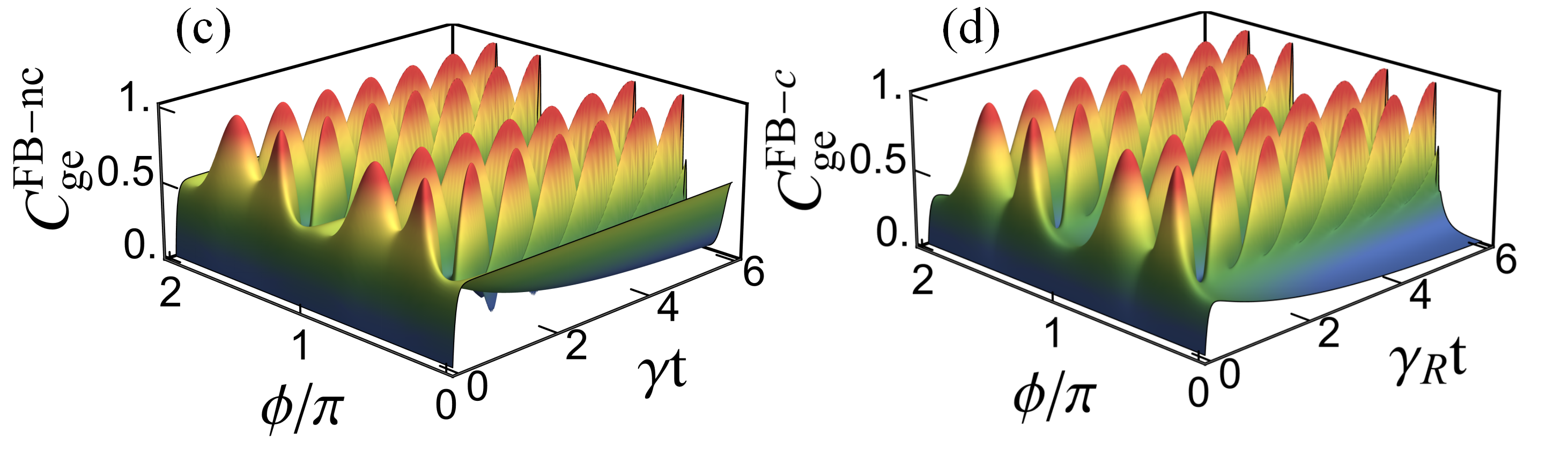}
	\includegraphics[width=0.48\textwidth]{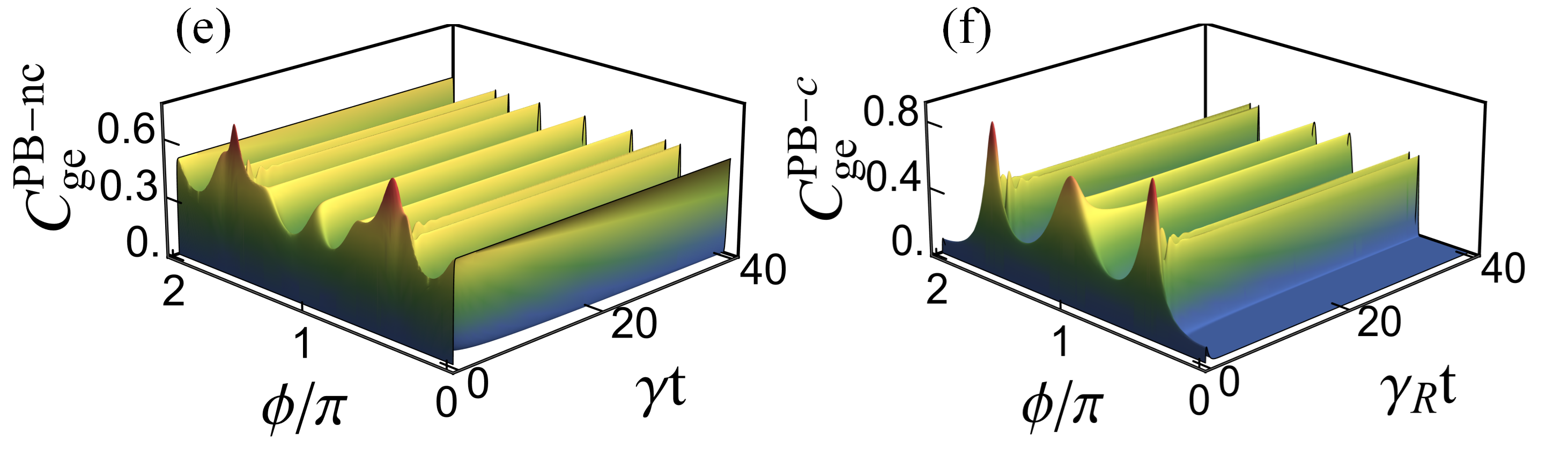}
	\includegraphics[width=0.48\textwidth]{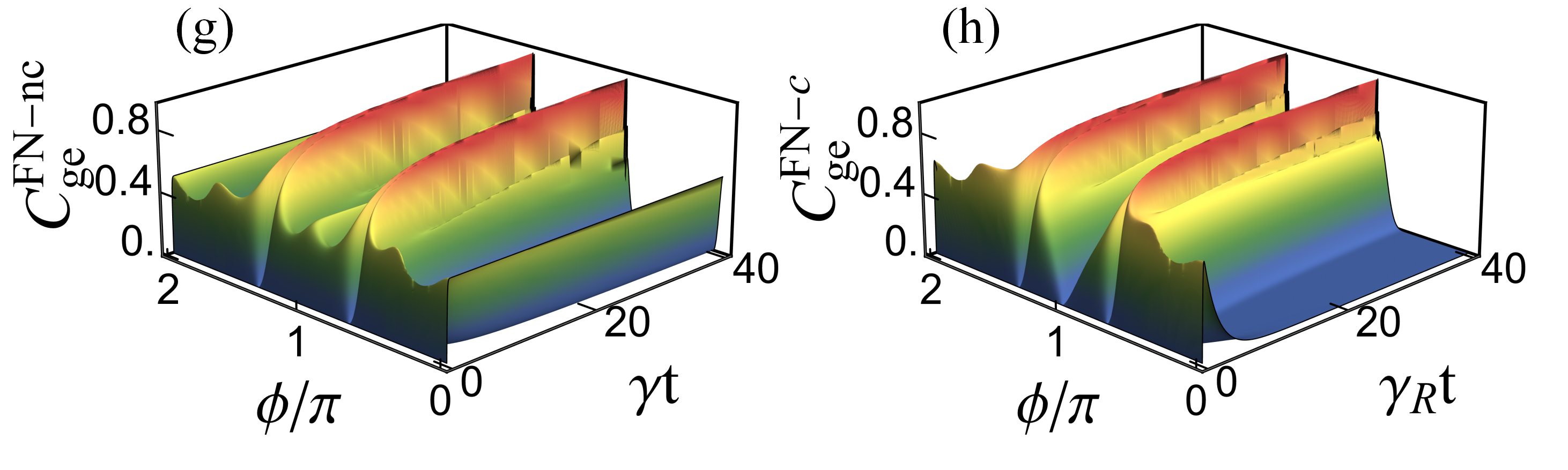}
	\includegraphics[width=0.48\textwidth]{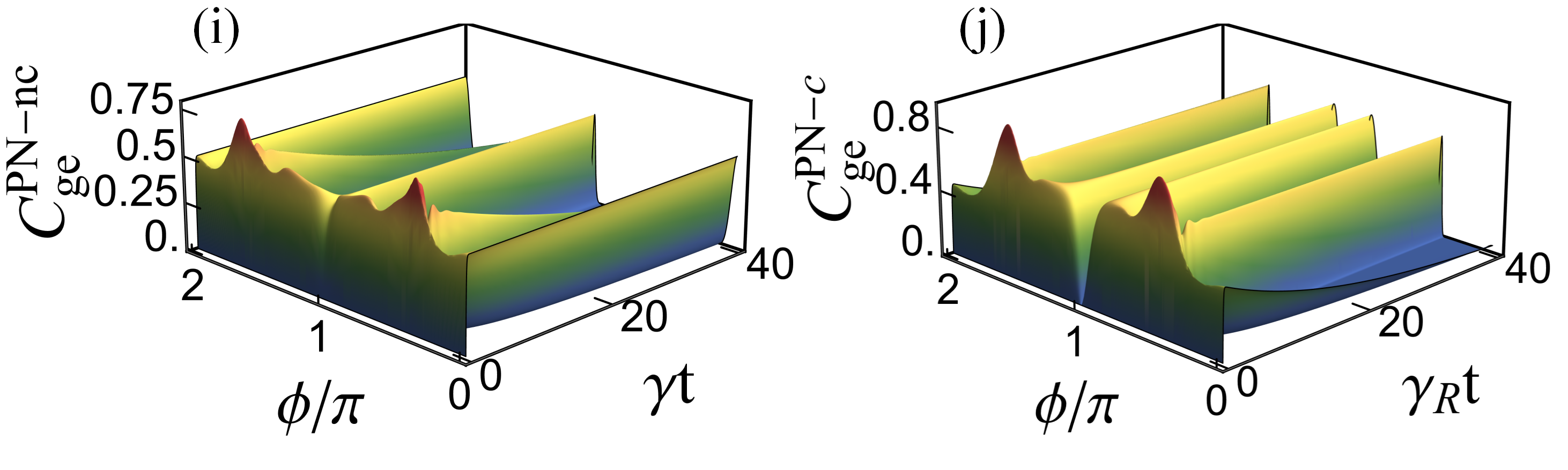}
	\caption{(Color online) Concurrence against $\gamma t$ ($\gamma_R t$) and $\phi/\pi$for the five configurations shown in Fig. \ref{fig1}. The giant atoms are initially in the state $|g_a,e_b\rangle $.}
	\label{fig7}
\end{figure}

In Sec. \ref{III}, we focus on the entanglement properties of two giant atoms in the single-excitation initial state $|e_a,g_b\rangle $. In order to investigate the impact of the atomic initial state on the entanglement, we now explore the cases when the initial state of the two giant atoms is set to $|g_a,e_b\rangle $ for five configurations shown in Fig. \ref{fig1}. The numerical results are illustrated in Fig. \ref{fig7}.

In the Figs. \ref{fig7}($a$), \ref{fig7}($c$) and \ref{fig7}($e$), we can directly observe that the evolution of concurrences $C_{ge}^{S-nc}$, $C_{ge}^{FB-nc}$ and $C_{ge}^{PB-nc}$, as functions of $\gamma t$ and $\phi/\pi$ in the separated and two braided configurations, remains consistent with the ones in Fig. \ref{fig2}($a_{1}$), Fig. \ref{fig3}($a_{1}$) and Fig. \ref{fig4}($a_{1}$). This is because the two giant atoms are equivalent in the separated and two braided configurations: having equal frequency shifts $\delta\omega_a$ = $\delta\omega_b$ and individual decay rates $\Gamma_a$ = $\Gamma_b$, satisfy the permutation symmetry. Consequently, these configurations exhibit the same entanglement dynamics, regardless of whether the initial state of the two giant atoms to be $|e_a,g_b\rangle $ or $|g_a,e_b\rangle $. Therefore, for two equivalent giant atoms in the single-excitation subspace under nonchiral coupling, the expression for concurrence satisfies
\begin{equation}
	C_{eg}= C_{ge}.\label{concurrence}
\end{equation}
However, the two nested configurations exhibit different features in entanglement dynamics under two different initial states, by comparing Fig. \ref{fig5}($a_{1}$) and  Fig. \ref{fig6}($a_{1}$) with Figs. \ref{fig7}(g) and  \ref{fig7}(i). By observing the two giant atoms in Figs. \ref{fig1}(d) and \ref{fig1}(e), we find that they are of different sizes (i.e., they are not equivalent) and do not satisfy permutation symmetry, leading to distinct entanglement evolution for the two initial states.

When giant atoms and waveguide are chirally coupled, the choice of initial state will affect the generation and evolution of entanglement, without regard to whether the two giant atoms are equivalent. Next, we will detail the reasons for the differences in entanglement evolution under different initial states from both analytical expressions and physical mechanisms. 

We begin our analysis by analyzing expressions. For the separated configuration, as shown in Fig. \ref{fig7}(b), we see that concurrence $C_{ge}^{S-c}$ is independent of $\phi$ and always maintains its initial value $C_{ge}^{S-c}$= 0, which is very different from the concurrence $C_{eg}^{S-c}$ in Fig. \ref{fig2}($a_{2}$) when the initial state is $|e_a,g_b\rangle $ for two giant atoms. To explain this phenomenon, we begin our analysis with Eq. (\ref{Concurrence}). When we upon substituting the parameters $g$, $\Gamma_j$, $\delta\omega_j$ and $\Gamma_{\text{coll}}$ of the giant atoms in this configuration into the expression for Eq. (\ref{Concurrence}), we find that $c_{eg}(t) = 0$, implying that the probability of the two giant atoms being in the state $|e_a,g_b\rangle $ is 0. Thus, with chiral coupling, we obtain a separated state $|g_a,e_b\rangle $ instead of an entangled state in this configuration. 
For the fully braided configuration, as observed in Fig. \ref{fig7}(d), the concurrence between two braided giant atoms shows oscillations ranging from 0 to 1 at phase shifts of $\phi = \pi/3$ and $2\pi/3$, consistent with the ones in Fig. \ref{fig3}($a_{2}$). This shows that the interatomic DF interactions still exist in the initial state $|g_a,e_b\rangle$ at these two phase shifts. However, for the phase shift $\phi =0$, $\pi$ and $2\pi$, the value of concurrence $C_{ge}^{FB-c}$ decreases compared to $C_{eg}^{FB-c}$. In this configuration, although the two giant atoms are equivalent, by comparing Fig. \ref{fig3}($a_{2}$) and Fig. \ref{fig7}(d), we find that the evolution of concurrences $C_{eg}^{FB-c}$ and $C_{ge}^{FB-c}$ is not identical. Which is due to the reason that the strength of the interatomic interactions $g$ and the collective dissipation $\Gamma_{\text{coll}}$ in the chiral case are not always real. Hence, according to Eq. (\ref{Concurrence}), the product $c_{eg}(t) c_{ge}^{\ast }(t)$ under the initial state $|e_a,g_b\rangle $ not be always identical to $c_{eg}(t) c_{ge}^{\ast }(t)$ under the initial state $|g_a,e_b\rangle $. Thus, the concurrences of two equivalent giant atoms, $C_{eg}^{FB-c}$ and $C_{ge}^{FB-c}$, in the chiral case can exhibit different characteristics. 

Similarly, in the partially braided configuration shown in Fig. \ref{fig7}(f), it can also be observed that the dynamical evolution of concurrence $C_{ge}^{PB-c}$ does not exactly coincide with the concurrence $C_{eg}^{PB-c}$ depicted in Fig. \ref{fig4}($a_{2}$). This difference shares the same reason as in the fully braided configuration. 
Finally, in the two nested configurations, illustrated in Figs. \ref{fig7}(h) and \ref{fig7}(j), the concurrences $C_{ge}^{FN-c}$ and $C_{ge}^{PN-c}$ exhibits markedly distinct features from concurrences $C_{eg}^{FN-c}$ and $C_{eg}^{PN-c}$ shown in Fig. \ref{fig5}($a_{2}$) and Fig. \ref{fig6}($a_{2}$) when considering two initial states. Beyond the reasons mentioned within the fully and partially braided configurations, an additional factor causing this variation is the distinct size of the two giant atoms. Hence, with both of two factors working together, it is reasonable to expect that concurrence exhibits distinct features for different initial states in the chiral case.

In terms of the physical mechanism, in chiral coupling, the interaction between giant atoms and waveguides is significantly directional, leading to a direction-dependent path of photon propagation from one atom to another. From the above, it is clear that the present study focuses on the case where giant atoms are coupled only to modes propagating to the right in the waveguide. In the separated configuration, as shown in Fig. \ref{fig1}(a), when two giant atoms are initially in the single-excitation state $|e_a,g_b\rangle $, photons from giant atom $a$ can propagate to the right via the path $X_{a1} \rightarrow X_{a2} \rightarrow X_{a3} \rightarrow X_{b1} \rightarrow X_{b2} \rightarrow X_{b3}$, resulting in giant atom $b$ absorbing the photon and transitioning into the excited state, thereby altering the state of the two giant atoms to $|g_a,e_b\rangle $. Through this process, a coherent superposition state of the initial $|e_a,g_b\rangle $ and final $|g_a,e_b\rangle $ states can be formed, i.e. entangled state. However, when the single-excitation initial state of the two giant atoms is $|g_a,e_b\rangle $, due to the directionality of the chiral coupling, the photons released by the giant atom $b$ can only propagate to the right along the waveguide, and cannot propagate to the left to be absorbed by the giant atom $a$. Therefore, two giant atoms cannot be in the single-excitation state $|e_a,g_b\rangle $ at any moment, and in view of the mechanism of the entanglement generation in the single-excitation subspace, at this time, there will not be between two giant atoms in this configuration entanglement generation.
For the other four configurations in Fig. \ref{fig1}, entanglement is generated between two giant atoms in the single-excitation state $|g_a,e_b\rangle $ when chirality is considered. The physical mechanisms for the difference in the entanglement evolution under the two single-excitation initial states $|g_a,e_b\rangle $ and $|e_a,g_b\rangle $ in the chiral coupling case are similar. Here, we will use the fully braided configuration shown in Fig. \ref{fig1}(b) as an example to explain the reason for the difference. As shown in Fig. \ref{fig1}(b), when the single-excitation initial state of the two giant atoms is $|e_a,g_b\rangle $, the propagation path of photons in giant atom $a$ is $X_{a1} \rightarrow X_{b1} \rightarrow X_{a2} \rightarrow X_{b2} \rightarrow X_{a3} \rightarrow X_{b3}$. Conversely, with the atoms in the state $|g_a,e_b\rangle$ at initial moment, the propagation path of photons in giant atom $b$ is $X_{b1} \rightarrow X_{a2} \rightarrow X_{b2} \rightarrow X_{a3} \rightarrow X_{b3}$. Therefore, different initial states lead to photon propagation along different paths, which affects the generation and evolution of entanglement. This finding emphasizes the importance of initial quantum states in controlling the entanglement dynamics of giant atomic systems, especially when chiral effects are considered.

\begin{table} 
	\caption{Maximum entanglement \( C_{\text{max}} \) generation in five setups when two giant atoms are coupled to the bidirectional-chiral waveguide. The first and second entries in this field are respective representations of the states \( |e_a,g_b\rangle \) and \( |g_a,e_b\rangle \).}
	\centering
	\begin{tabularx}{\textwidth}{@{}l*{5}{>{\centering\arraybackslash}X}@{}} 
		\toprule
		\textbf{} & \textbf{separated} & \textbf{fully braided} & \textbf{partially braided} & \textbf{fully nested} & \textbf{partially nested} \\
		\midrule
		nonchiral coupling & 0.5; 0.5 & 1; 1 & 0.77; 0.77 & 0.87; 0.96 & 0.83; 0.78 \\
		chiral coupling & 0.736; 0 & 1; 1 & 0.86; 0.89 & 0.90; 0.98 & 0.94; 0.93 \\
		\bottomrule
	\end{tabularx}
	\label{tab:Example}
\end{table}

In order to see the maximum entanglement under two different initial states $|e_a,g_b\rangle $ and $|g_a,e_b\rangle $ more intuitively and clearly, we list the maximum entanglement $C_{max}$ for five configurations in Table \ref{tab:Example}. From Table \ref{tab:Example}, we can directly see that in the fully braided configuration, because of the presence of DF interaction, its maximum entanglement value $C_{max}$ can reach 1 in both states $|e_a,g_b\rangle $ and $|g_a,e_b\rangle $. Hence, the fully braided configuration represents our optimal choice for preparing entangled states within the single-excitation subspace.

\section{THE ROLE OF IMPERFECT CHIRALITY IN FULLY BRAIDED GIANT ATOM CONFIGURATION}\label{V}

In the above contents, we have considered the generation of entanglement between two giant atoms in a bidirectional-chiral waveguide across various configurations, within a single-excitation subspace. Among these, the two giant atoms in the fully braided configuration, irrespective of being nonchiral or chiral, has shown a remarkable characteristic: two atoms along the waveguide are capable of interacting without photon emissions from the giant atoms. That is, the two atoms have DF interactions. This unique property enables the maximum entanglement value between two giant atoms can reach 1, a feature not observed in small atom system \cite{Gonzalez-Ballestero40} and other configurations discussed in this paper.

Our previous findings were predicated on the assumption of perfect chirality, an idealized case that may not always be feasible. Given the significant role of the fully braided configuration in maximizing entanglement, it is crucial to examine its behavior under more realistic conditions, particularly considering the presence of imperfect chirality which is more likely in actual physical systems. Therefore, we have devoted this section to specifically explore the impact of imperfect chirality on entanglement maximum value in the fully braided configuration. By focusing on this configuration under conditions of imperfect chirality, we aim to understand how deviations of real physical systems from the ideal case influence the dynamics of entanglement. Here, the chosen initial state for the two giant atoms is $|e_a,g_b\rangle $. 

\begin{figure}[tbp]
	\center\includegraphics[width=0.98\textwidth]{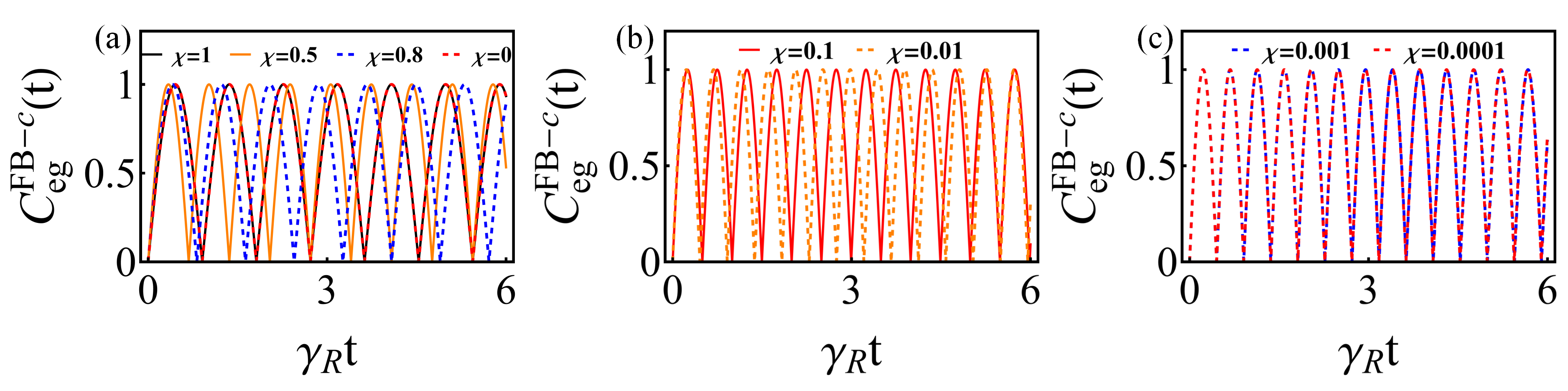}
	\caption{(Color online) $(a)$ Time evolution of concurrence $C_{\text{eg}}^{FB-c}(t)$ with different chirality $\chi$. $(b)$ Time evolution of concurrence $C_{\text{eg}}^{FB-c}(t)$ for $\chi$=0.1, 0.01. $(c)$ Time evolution of concurrence $C_{\text{eg}}^{FB-c}(t)$ for $\chi$=0.001, 0.0001. In $(a)$, $(b)$ and $(c)$, the phase shift $\phi$= $\pi/3$ or $2\pi/3$.}
	\label{fig8}
\end{figure}

To accomplish this, we define chirality as $\chi = (\gamma_R - \gamma_L) / (\gamma_R + \gamma_L)$, with values ranging from $0$ to $1$. Here, $0$ corresponds to the bidirectional case (i.e., nonchiral case), while $1$ corresponds to perfect chirality. The time evolution of the concurrence $C_{\text{eg}}^{FB-c}(t)$ is illustrated in Fig. \ref{fig8}(a) for specific phase shifts when $\chi$ takes different values. One can observe that the maximum entanglement value of concurrence $C_{\text{eg}}^{FB-c}(t)$ is independent of $\chi$ and consistently remains at 1, i.e., the maximum entanglement value is robust to $\chi$, which is not possible in small-atom chiral system \cite{Gonzalez-Ballestero40}. This is primarily attributed to the fact that chirality does not influence the DF interaction \cite{Soro51, Carollo}, guaranteeing the maximum entanglement value immune to variations of $\chi$. Remarkably, a counterintuitive phenomenon emerges: when setting chirality $\chi$ to both $0$ and $1$, the evolution of concurrence completely overlaps, as depicted by the black solid and red dashed curves in Fig. \ref{fig8}(a). This intriguing behavior sets this configuration apart from others and small atomic cases. To clarify this observation, we examine the the values of parameters for the two fully braided giant atoms at $\chi = 0$ and $\chi = 1$. Our calculations show that collective dissipation $\Gamma_{\text{coll}}$ are all zero, and the value of coupling strength $g$ are precisely the same in both cases. Thus, the concurrence $C_{\text{eg}}^{FB-c}(t)$ exhibits same dynamical evolution for both $\chi$ = 0 and $\chi$ = 1. For other chiralities $\chi$ in Fig. \ref{fig8}(a), dissipative $\Gamma_{\text{coll}}$ are still zero, while the non-zero coupling strength $g$ varies in magnitude, resulting in distinct evolutions of concurrence $C_{\text{eg}}^{FB-c}(t)$. 
Additionally, Fig. \ref{fig8}(a) reveals that smaller $\chi$ values (except for $\chi$=0) correspond to more occurrences of the peak value $C_{\text{eg}}^{FB-c}(t) = 1$ within the same timescale. Based on this observation, we investigated how the occurrence of peak values is influenced by variations in $\chi$, by setting the chirality parameter $\chi$ to smaller non-zero values and adjusting the phase shift between two coupling points. For this purpose, we plotted concurrence $C_{\text{eg}}^{FB-c}(t)$ as a function of $\gamma_R t$ for various chirality $\chi = 0.1, 0.01, 0.001, 0.0001$, as depicted in Figs. \ref{fig8}(b) and \ref{fig8}(c). It can be easily observed that as the chirality $\chi$ is reduced from 0.1 to 0.01, the number at which $C_{\text{eg}}^{FB-c}(t)=1$ occurs slightly increases within the considered timescale, as illustrated in Fig. \ref{fig8}(b). With further reduction of $\chi$, the curves show no significant changes, as demonstrated in Fig. \ref{fig8}(c). Therefore, these results suggest that while the initial reduction of chirality $\chi$ leads to an increase in the number of achieving the peak value 1, further reductions in $\chi$ have a limited impact on this increment. This discovery highlights the importance of precisely controlling the chirality parameter $\chi$ when optimizing system parameters to enhance performance.
\section{CONCLUSION}\label{VI}

In summary, we explore the generation of entanglement of two giant atoms under nonchiral and chiral couplings in a giant-atom waveguide-QED system. We mainly focus on the entanglement generation of two giant atoms initially in the single-excitation states $|e_a,g_b\rangle $ in five configurations. Our study shows that the entanglement, influenced by phase shift and coupling configurations, can exhibit oscillatory decay or steady state characteristics. In nonchiral case, due to the presence of dark state, we can obtain steady-state entanglement in all five configurations. Comparing with the work of Yin \emph{et al.} \cite{Yin55}, we find that, in the four configurations other than the fully braided one, increasing the number of coupling points can generate steady-state entanglement at more phase shifts. However, the chiral giant-atom waveguide-QED system, in comparison to nonchiral giant atom and chiral small atom systems, can maximally enhance entanglement. Furthermore, we demonstrate that in fully braided configuration, the maximum entanglement value can reach 1, which remains insensitive to chirality, a feature not achievable in small-atom chiral quantum networks \cite{Gonzalez-Ballestero40}. We further investigated how the initial states of atoms influence the entanglement dynamics between two giant atoms. Our results reveal that, for different single-excitation initial states, under chiral coupling condition, variations in entanglement dynamics occur regardless of whether the two giant atoms exhibit permutation symmetry. Conversely, under nonchiral coupling case, differences in entanglement dynamics are observed only when the two giant atoms do not exhibit permutation symmetry.

Our schemes can be realized in superconducting circuits. Advances in experiments have successfully realized braided configurations of two artificial giant atoms in a cryogenic temperature superconducting qubit system with nonchiral coupling \cite{Kannan48}. Recently, chiral giant atoms with near-perfect directionality have also been realized experimentally \cite{Joshi63}. The study of ours provides a fundamental framework for the generation of entanglement within chiral giant-atom waveguide-QED system.

\section{ACKNOWLEDGMENTS}
This work was supported by National Natural Science Foundation of China(Grants No. 11874190 and No. 12247101). Support was also provided by Supercomputing Center of Lanzhou University.
\bibliography{REF}
\end{document}